\documentclass[letterpaper,11pt]{article}

\usepackage{fancyhdr}
\usepackage{amsmath}
\usepackage{amsbsy}
\usepackage{amssymb}
\usepackage{amsthm}
\usepackage{amscd}
\usepackage{amsfonts}
\usepackage{supertabular}
\usepackage{graphics}
\usepackage{graphicx}
\usepackage{verbatim}
\usepackage{subfigure}
\usepackage{epsfig}
\usepackage{xspace}
\usepackage{euscript}
\usepackage{alltt}
\usepackage{boxedminipage}
\usepackage{float}
\usepackage{times}
\usepackage[colorlinks]{hyperref}
\usepackage[square,numbers]{natbib}
\usepackage{setspace}
\usepackage{authblk}

\usepackage{algorithm}
\usepackage{algorithmic}

\def\Put(#1,#2)#3{\leavevmode\makebox(0,0){\put(#1,#2){#3}}}

\newcommand{\qvec}[1]{{\boldsymbol{#1}}}

\begin{document}

\title{Microstructure evolution of compressible granular systems under large deformations}

\author[1]{Marcial Gonzalez \thanks{marcial-gonzalez@purdue.edu}}
\affil[1]{School of Mechanical Engineering, Purdue University, \break West Lafayette, IN 47907, USA}
\author[2]{Alberto M. Cuiti\~{n}o \thanks{cuitino@jove.rutgers.edu}}
\affil[2]{Department of Mechanical and Aerospace Engineering, Rutgers University, \break Piscataway, NJ 08854, USA}

\maketitle

\begin{abstract}
We report three-dimensional particle mechanics static calculations that predict the microstructure evolution during die-compaction of elastic spherical particles up to relative densities close to one. We employ a nonlocal contact formulation that remains predictive at high levels of confinement by removing the classical assumption that contacts between particles are formulated locally as independent pair-interactions. The approach demonstrates that the coordination number depends on the level of compressibility, i.e., on the Poisson's ratio, of the particles. Results also reveal that distributions of contact forces between particles and between particles and walls, although similar at jamming onset, are very different at full compaction. Particle-wall forces are in remarkable agreement with experimental measurements reported in the literature, providing a unifying framework for bridging experimental boundary observations with bulk behavior.
\end{abstract}

%%%%%%%%%%%%%%%%%%%%%%%%%%%%

\section{Introduction}

The microstructure of confined granular media is typically inhomogeneous, anisotropic and disordered. Under external loading, these systems exhibit a non-equilibrium jamming transition from a liquid-like to a solid-like state \cite{Liu-1998,Jin-2010,Trappe-2001,OHern-2003,Song-2008}. Under increasing confinement, these amorphous solids support stress by spatial rearrangement and deformation of particles and by the development of inhomogeneous force networks \cite{Majmudar-2005,Durian-1995,Makse-2000,Blair-2001,Liu-1995,Radjai-1996,Chan-2005}. Understanding and predicting the formation and evolution of such microstructure under finite macroscopic deformations and at high levels of confinement remains a fundamental goal of granular mechanics \cite{Edwards-1989,Cates-1998,Richard-2005,Blumenfeld-2009,Ciamarra-2012,Ostojic-2006}.

In this paper, we report three-dimensional particle mechanics static calculations that enable us to predict microstructure evolution during die-compaction of elastic spherical particles up to relative densities close to one. We employ a nonlocal contact formulation \cite{Gonzalez-2012} that remains predictive at high levels of confinement by removing the classical assumption that contacts between particles are formulated locally as independent pair-interactions. We specifically study a noncohesive frictionless granular system comprised by 40,000 weightless elastic spherical particles with radius $R=0.125$~mm, Young's modulus $E=7.0$~GPa, and Poisson's ratio $\nu$. The granular bed, which is numerically generated by means of a ballistic deposition technique \cite{Jullien-1989}, is constrained by a rigid cylindrical container of diameter $10$~mm. The systems is deformed under die-compaction up to a relative density $\rho$ of $1$. Assuming a sufficiently small compaction speed, we consider rate-independent material behavior and we neglect traveling waves, or any other dynamic effect \cite{Richard-2005}. The deformation process is therefore described by a sequence of static equilibrium configurations. In this work we employ 125 quasi-static load steps and we consider two materials that only differ in their level of compressibility, namely $\nu=0.35$ and $\nu=0.45$.

The paper is organized as follows. The particle mechanics approach used to generate a sequence of static equilibrium configurations of granular systems at high levels of confinement is presented in Section \ref{Section-ParticleMechanics}. The evolution of microstructural statistical features of these equilibrium configurations is investigated in Section \ref{Section-PostJamming}. Specifically, we study the evolution of the mechanical coordination number (number of non-zero contact forces between a particle and its neighbors) in section \ref{Subsection-Coordination}, punch and die-wall pressures in section \ref{Subsection-Pressure}, the network of contact forces in section \ref{Subsection-NetworkCF}, and the network of contact radiuses in section \ref{Subsection-NetworkCR}. Finally, a summary and concluding remarks are collected in Section \ref{Section-Summary}.

\section{Particle mechanics approach to granular systems at high levels of confinement}
\label{Section-ParticleMechanics}

\begin{figure}[htbp]
\centering{
    \includegraphics[scale=0.64]{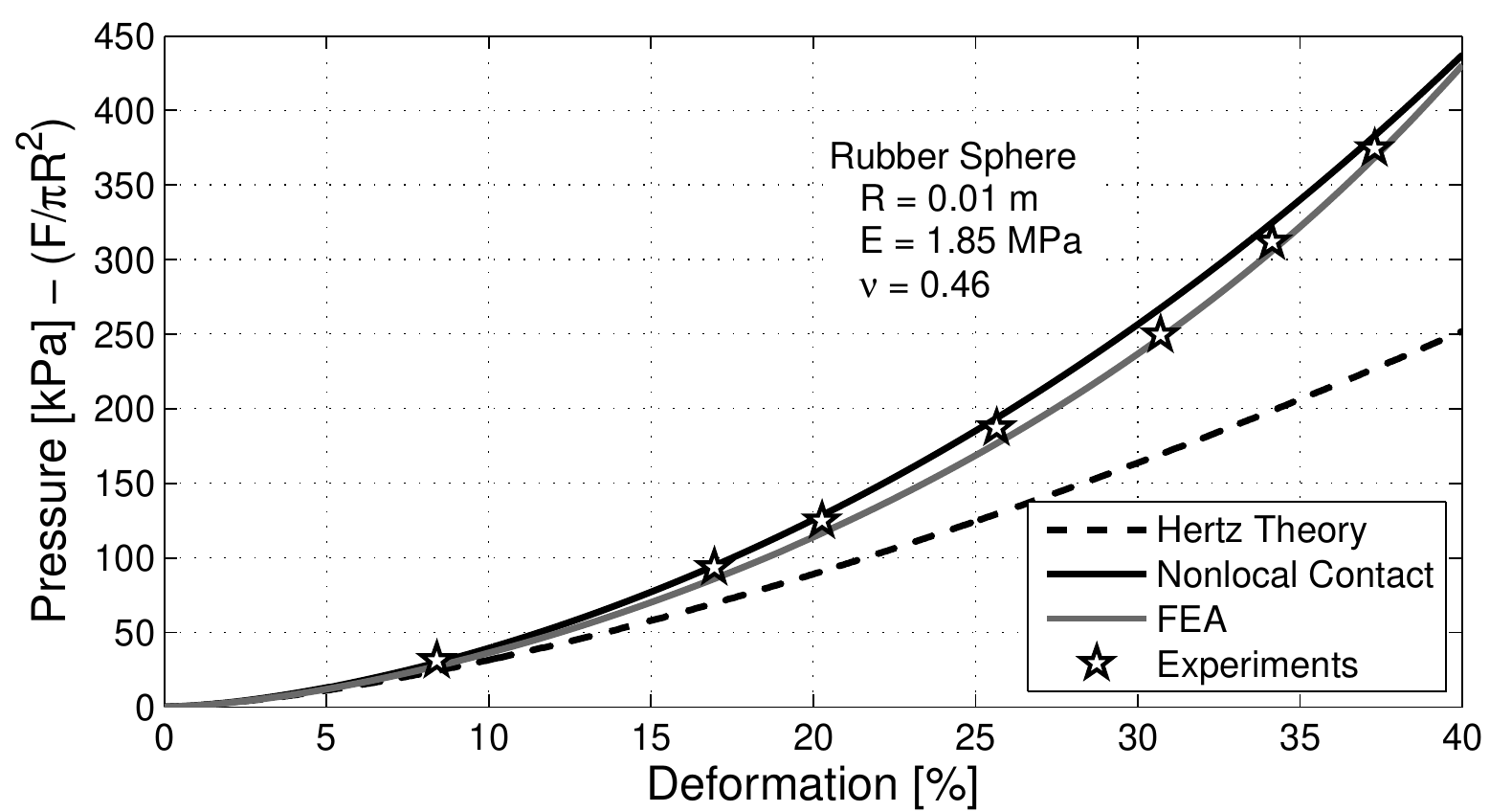}
    \Put(-240,212){\includegraphics[scale=0.60]{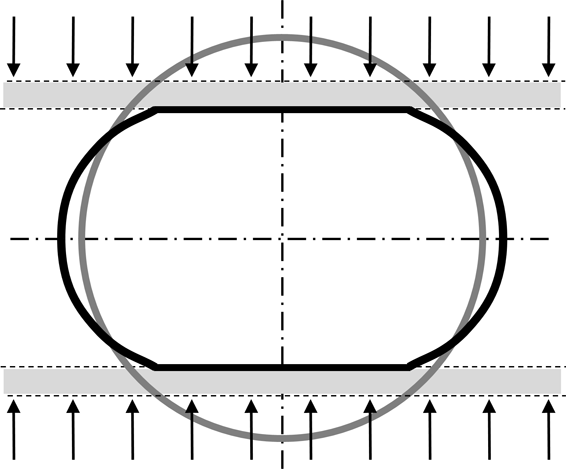}}
}
\caption{Validation of the nonlocal contact formulation with experimental measurements obtained for a rubber sphere compressed between two rigid plates \cite{Tatara-1989} and with detailed finite-element simulations  \cite{Gonzalez-2012}.}
\label{Fig-1D-Experiment}
\end{figure}

Due to the high level of confinement experienced by the particles in the system, we adopt a nonlocal contact formulation \cite{Gonzalez-2012} that accounts for the interplay of deformations due to multiple contact forces acting on each single particle. This nonlocal formulation removes the classical assumption that contacts between particles are  formulated locally as independent pair-interactions. Its predictions are in remarkable agreement with detailed finite-element simulations of a neo-Hookean solid extended to the compressible range and with experimental observations of a rubber sphere that exhibits no permanent deformations. This good agreement at moderate to large levels of confinement, however, is not obtained with Hertz theory (see Figure~\ref{Fig-1D-Experiment} and \cite{Gonzalez-2012} for further details). It bears emphasis that the nonlocal contact formulation is not an extension of Hertz theory to finite deformations but rather a systematic approach for accounting the interaction between contact interfaces. In a manner analogous to interacting cracks or dislocations in elastic media, the nonlinear behavior is confined to small regions and the interaction between these regions is through an elastic field of deformations.

\begin{figure}[htbp]
\centering{
    \includegraphics[scale=0.55]{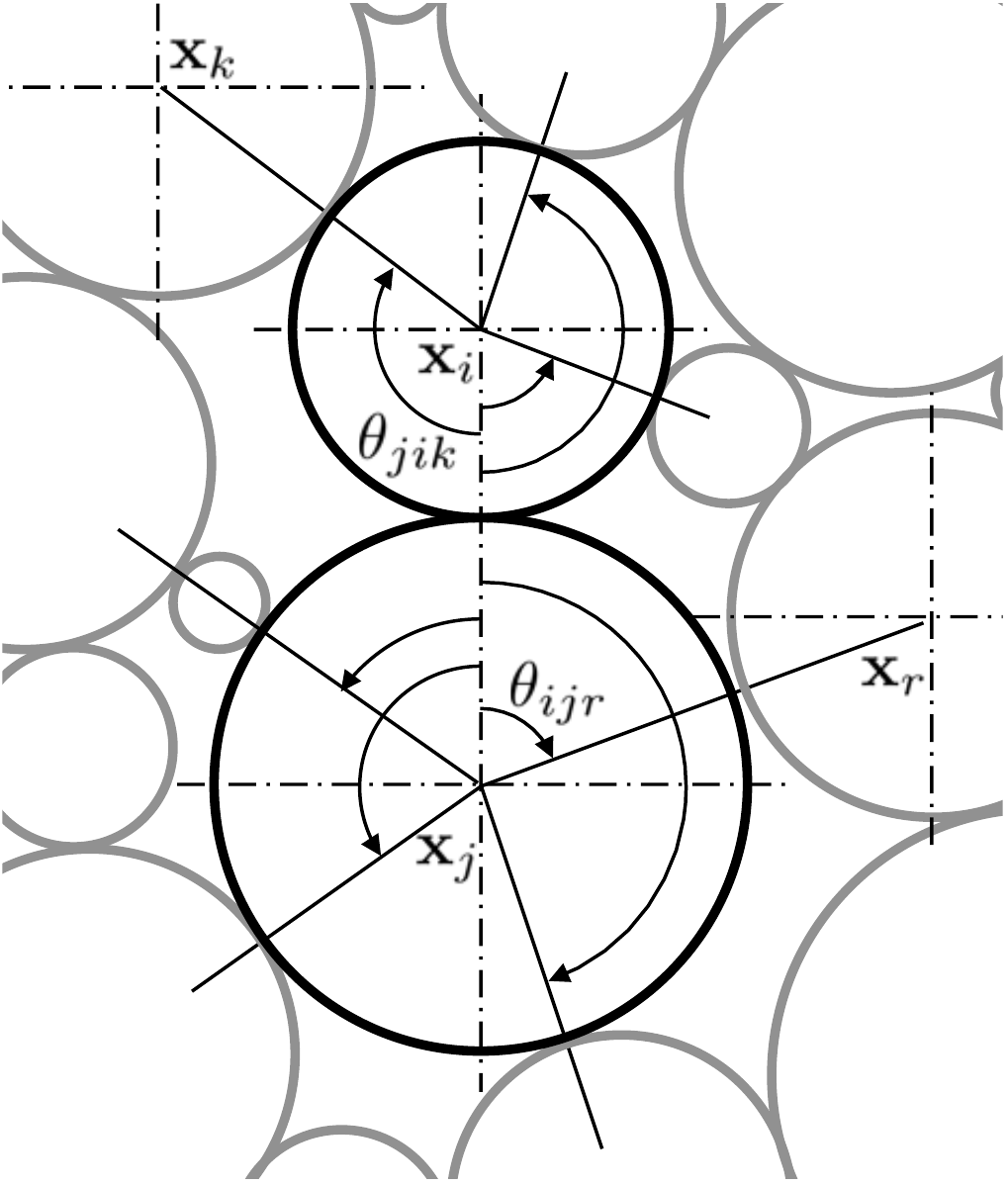}
}
\caption{Two spheres in contact under a general loading configuration comprised by multiple neighbors.}
\label{Fig-NonlocalAngles}
\end{figure}

An equilibrium configuration is defined by the solution of two sets of coupled nonlinear equations. The first set of equations corresponds to static equilibrium of the granular system, that is sum of all elastic contact forces acting on each particle equals zero. The second set of equations accounts for the contribution to each contact interface of the nonlocal mesoscopic deformations induced by all other contact forces acting on the particles, that is for the nonlocal terms $\gamma^{\textrm{\tiny{NL}}}$ \cite{Gonzalez-2012}. Thus, the nonlinear system of equations is given by
\begin{eqnarray}
\label{Eqn-SOE}
    \sum\nolimits_{j \in \mathcal{N}_i} n^{\textrm{\tiny{H}}}_{ij}
                              (R_i+R_j-\|\mathbf{x}_i - \mathbf{x}_j\| + \gamma^{\textrm{\tiny{NL}}}_{ij})_+^{3/2}
                              \tfrac{\mathbf{x}_i - \mathbf{x}_j}{\|\mathbf{x}_i - \mathbf{x}_j\|}
    = &\mathbf{0} \nonumber
    \\
    \gamma^{\textrm{\tiny{NL}}}_{ij}
    -
    \sum\nolimits_{k \in \mathcal{N}_i,k \neq j}
    \tfrac{n^{\textrm{\tiny{H}}}_{ik} (R_i+R_k-\|\mathbf{x}_i - \mathbf{x}_k\| + \gamma^{\textrm{\tiny{NL}}}_{ik})_+^{3/2}}{n^{\textrm{\tiny{NL}}}_{jik}(\mathbf{x}_j,\mathbf{x}_i,\mathbf{x}_k)}
    -\hspace{.35in}&  \\*
    \sum\nolimits_{k \in \mathcal{N}_j,k \neq i}
    \tfrac{n^{\textrm{\tiny{H}}}_{jk} (R_j+R_k-\|\mathbf{x}_j - \mathbf{x}_k\| + \gamma^{\textrm{\tiny{NL}}}_{jk})_+^{3/2}}{n^{\textrm{\tiny{NL}}}_{ijk}(\mathbf{x}_i,\mathbf{x}_j,\mathbf{x}_k)}
    = &0 \nonumber
\end{eqnarray}
where $\mathbf{x}_i$ and $\mathcal{N}_i$ are the position and all the neighbors of particle $i$, respectively, $\gamma^{\textrm{\tiny{NL}}}_{ij} = \gamma^{\textrm{\tiny{NL}}}_{ji}$ by definition, and $(\cdot)_+=\max\{\cdot,0\}$. In the above system of equations, $n^{\textrm{\tiny{H}}}_{ij}$ and $n^{\textrm{\tiny{NL}}}_{jik}$ correspond to Hertz theory and to the nonlocal contact formulation \cite{Gonzalez-2012}, respectively, with
\begin{eqnarray}
    n^{\textrm{\tiny{H}}}_{ij}
    &=
    &\tfrac{4}{3}
    \left( \tfrac{1-\nu_i^2}{E_i} + \tfrac{1-\nu_j^2}{E_j} \right)^{-1}
    \left( \tfrac{1}{R_i} + \tfrac{1}{R_j} \right)^{-1/2}
    \\
    n^{\textrm{\tiny{NL}}}_{jik}
    &=
    &\tfrac{4\pi R_i E_i\sin(\theta_{jik}/2)}
          {(1+\nu_i) [-2(1-\nu_i)-2(1-2\nu_i)\sin(\theta_{jik}/2)+(7-8\nu_i)\sin^2(\theta_{jik}/2)]} \nonumber
\end{eqnarray}
where $\theta_{jik}=\widehat{x_jx_ix_k}$ is the angle between the coordinate points of particles $j$, $i$ and $k$ (see Figure~\ref{Fig-NonlocalAngles}).

A sequential strategy is proposed to treat the coupled problem (see Algorithm \ref{Alg-SequentialStrategy}). The equations of static equilibrium are solved for $\mathbf{x}=(\mathbf{x}_1^T, ..., \mathbf{x}_N^T)^T$, assuming fixed $\qvec{\gamma}^{\textrm{\tiny{NL}}}=(\gamma_{12}^{\textrm{\tiny{NL}}}, \gamma_{13}^{\textrm{\tiny{NL}}}, ...,\gamma_{N-1,N}^{\textrm{\tiny{NL}}})^T$, by employing a trust-region method \cite{Coleman-1996,Conn-2000} that successfully overcomes the characteristic ill-posedness of the problem (e.g., due to metastability \cite{Mehta-2007}). The basic trust-region algorithm requires the solution of a minimization problem to determine the step between iterations, namely the trust-region step. This minimization problem is of the form $\min\{\psi(\mathbf{s}):\|\mathbf{s}\| \leq \Delta\}$, where $\mathbf{s}={^{n+1}\mathbf{x}}-{^{n}\mathbf{x}}$ is the trust-region step, $\Delta$ is a trust-region radius and $\psi$ is a quadratic function that represents a local model of the objective function about ${^{n}\mathbf{x}}$, that is 
$$
\psi(\mathbf{s})= \| {^n}\mathbf{F} + {^n}\mathbf{K} \mathbf{s} \|^2 
= 
\tfrac{1}{2} \left< {^n}\mathbf{F},{^n}\mathbf{F} \right> + \left< {^n}\mathbf{K} \mathbf{s}, {^n}\mathbf{F} \right> + \tfrac{1}{2} \left< {^n}\mathbf{K} \mathbf{s}, {^n}\mathbf{K} \mathbf{s}\right>
$$ 
with ${^n}\mathbf{F}$ and ${^n}\mathbf{K}$ the global force vector and stiffness matrix at ${^{n}\mathbf{x}}$---the first term in the above equation is not required in the minimization problem. It is worth noting that the trust-region step is not necessarily in the direction of a quasi-Newton step and that the trust-region radius acts as a regularization term that controls the growth in the size of the least squares solution observed in most ill-posed \cite{Vicente-1996}. Trading accuracy for performance, Byrd \cite{Byrd-1988}, among others, proposed to approximate the minimization problem by restricting the problem to a two-dimensional subspace. Furthermore, the two-dimensional subspace can be determined by a preconditioned conjugate gradient process and the trust-region radius can be adjusted over the iterative process (see, e.g., \cite{More-1983}). Here we adopt the implementation available in MATLAB R2014a Optimization Toolbox.

The nonlocal terms are solved by a few fixed point iterations of the second set of equations in (\ref{Eqn-SOE}), assuming a given $\mathbf{x}$ (see \cite{Gonzalez-2012} and Algorithm \ref{Alg-NonlocalAssembly}). It is worth noting that an equilibrium configuration is not obtained by artificially dumped or cooled-down dynamic processes but rather by iterative solvers that follow the energy landscape around the solution of static equilibrium. Each iteration of the trust-region method involves the solution of $\qvec{\gamma}^{\textrm{\tiny{NL}}}$ (subject to a given tolerance TOL2) and the evaluation of an approximate solution for $\mathbf{x}$. A notable feature of the nonlocal contact formulation is that, when nonlocal effects are neglected, it reduces to Hertz theory. Under such simplification, an equilibrium configuration is readily obtained from the solution of the first set of equations in (\ref{Eqn-SOE}), assuming $\qvec{\gamma}^{\textrm{\tiny{NL}}}=0$.

\begin{algorithm}[htbp]
\caption{ParticleMechanicsApproach: sequential strategy for solving the coupled equilibrium problem.}
\label{Alg-SequentialStrategy}
\begin{algorithmic}[1]
\REQUIRE Initial guess for particles' coordinates ${^1}\mathbf{x}$, initial guess for nonlocal terms ${^{0}}\qvec{\gamma}^\textrm{\tiny{NL}}$, TOL1, TOL2, trust-region radius $\Delta$ and the simplicial complex $X$
\STATE $\text{Error} \leftarrow \text{TOL1}$
\STATE $n \leftarrow 1$
\WHILE{$\text{Error} \geq \text{TOL1}$}
\STATE \textit{/$\ast$~~Compute global force, global stiffness, and nonlocal terms.~~$\ast$/}
\STATE $\{{^n}\mathbf{F},{^n}\mathbf{K},{^n}\qvec{\gamma}^\textrm{\tiny{NL}}\} \leftarrow \mathbf{NonlocaContactFormulation}({^n}\mathbf{x},\text{TOL2},{^{n-1}}\qvec{\gamma}^\textrm{\tiny{NL}},X)$
\STATE \textit{/$\ast$~~Update coordinates with the trust-region step obtained by restricting the problem to a two-dimensional subspace \cite{Byrd-1988}.~~$\ast$/}
\STATE ${^n\mathbf{s}} \leftarrow \text{argmin}_\mathbf{s} ~  \left< {^n}\mathbf{K} \mathbf{s}, {^n}\mathbf{F} \right> + \tfrac{1}{2} \left< {^n}\mathbf{K} \mathbf{s}, {^n}\mathbf{K} \mathbf{s}\right>$ ~~subject to~~ $\|\mathbf{s}\| \le \Delta$
\STATE ${^{n+1}\mathbf{x}} \leftarrow {^{n}\mathbf{x}} + {^n\mathbf{s}}$
\STATE \textit{/$\ast$~~Compute a measure of convergence~~$\ast$/}
\STATE $\text{Error} \leftarrow \| {^n\mathbf{s}} \|$
\STATE $n \leftarrow n+1$
\ENDWHILE
\STATE \textit{/$\ast$~~Compute nonlocal terms of the converged solution.~~$\ast$/}
\STATE $\{ {^n}\qvec{\gamma}^\textrm{\tiny{NL}} \} \leftarrow \mathbf{NonlocaContactFormulation}({^n}\mathbf{x},\text{TOL2},{^{n-1}}\qvec{\gamma}^\textrm{\tiny{NL}},X)$
\RETURN $\{ {^n\mathbf{x}}, {^n\qvec{\gamma}^\textrm{\tiny{NL}}} \}$ 
\end{algorithmic}
\end{algorithm}

\begin{algorithm}[htbp]
\caption{NonlocalContactFormulation: computation of global force, global stiffness, and nonlocal terms.}
\label{Alg-NonlocalAssembly}
\begin{algorithmic}[1]
\REQUIRE Particles' coordinates $\mathbf{x}=(\mathbf{x}_1^T, ..., \mathbf{x}_N^T)^T$, TOL, initial guess for nonlocal terms ${^0}\qvec{\gamma}^{\textrm{\tiny{NL}}}=({^0\gamma_{12}^{\textrm{\tiny{NL}}}}, {^0\gamma_{13}^{\textrm{\tiny{NL}}}}, ..., {^0\gamma_{N-1,N}^{\textrm{\tiny{NL}}}})^T$ and the simplicial complex $X$.
\STATE \textit{/$\ast$~~Compute nonlocal contributions $\qvec{\gamma}^{\textrm{\tiny{NL}}}$~~$\ast$/}
\STATE $\text{Error} \leftarrow \text{TOL}$
\STATE $n \leftarrow 0$
\WHILE{$\text{Error} \geq \text{TOL}$}
\STATE {$ ^{n+1}\qvec{\gamma}^{\textrm{\tiny{NL}}} \leftarrow \qvec{0}$}
\STATE \textit{/$\ast$~~Loop over all contact pair-interactions between vertices $\{ v_i, v_j\}$~~$\ast$/}
\FOR  {$[ v_i, v_j] \in E_1(X)$}
\FOR  {$ [v_i, v_{k\ne j}]  \in$ {\sc Incident}$\left( v_i \right)$}
\STATE \textit{/$\ast$~~Update $\gamma^{\textrm{\tiny{NL}}}_{ij}$ with the nonlocal displacement induced by contact force $P_{ik}$~~$\ast$/}
\STATE ${^{n+1}\gamma^{\textrm{\tiny{NL}}}_{ij} \leftarrow {^{n+1}\gamma^{\textrm{\tiny{NL}}}_{ij}} +\tfrac{n^{\textrm{\tiny{H}}}_{ik} (R_i+R_k-\|\mathbf{x}_i - \mathbf{x}_k\| + {^n}\gamma^{\textrm{\tiny{NL}}}_{ik})_+^{3/2}}{n^{\textrm{\tiny{NL}}}_{jik}(\mathbf{x}_j,\mathbf{x}_i,\mathbf{x}_k)}}$
\ENDFOR
\FOR  {$[v_j, v_{k\ne i}] \in$ {\sc Incident}$\left( v_j \right)$}
\STATE \textit{/$\ast$~~Update $\gamma^{\textrm{\tiny{NL}}}_{ij}$ with the nonlocal displacement induced by contact force $P_{jk}$~~$\ast$/}
\STATE ${^{n+1}\gamma^{\textrm{\tiny{NL}}}_{ij} \leftarrow {^{n+1}\gamma^{\textrm{\tiny{NL}}}_{ij}} +\tfrac{n^{\textrm{\tiny{H}}}_{jk} (R_j+R_k-\|\mathbf{x}_j - \mathbf{x}_k\| + {^n}\gamma^{\textrm{\tiny{NL}}}_{jk})_+^{3/2}}{n^{\textrm{\tiny{NL}}}_{ijk}(\mathbf{x}_i,\mathbf{x}_j,\mathbf{x}_k)}}$
\ENDFOR
\ENDFOR
\STATE \textit{/$\ast$~~Compute a measure of convergence~~$\ast$/}
\STATE $\text{Error} \leftarrow \| {^{n+1}\qvec{\gamma}^{\textrm{\tiny{NL}}}} - {^n\qvec{\gamma}^{\textrm{\tiny{NL}}}} \|$
\STATE $n \leftarrow n+1$
\ENDWHILE
\STATE \textit{/$\ast$~~Assemble global force $F$ and stiffness $K$ using converged nonlocal terms ${^n\qvec{\gamma}^{\textrm{\tiny{NL}}}}$~~$\ast$/}
\FOR  {$[v_i , v_j] \in E_1(X)$}
%\STATE $\mathbf{F}_{ij} \leftarrow n^{\textrm{\tiny{H}}}_{ij} (R_i+R_j-\|\mathbf{x}_i - \mathbf{x}_j\| + {^n}\gamma^{\textrm{\tiny{NL}}}_{ij})_+^{3/2} \tfrac{\mathbf{x}_i - \mathbf{x}_j}{\|\mathbf{x}_i - \mathbf{x}_j\|}$
\STATE $\gamma_{ij}  \leftarrow R_i+R_j-\|\mathbf{x}_i - \mathbf{x}_j\|$
\STATE $\mathbf{F} \leftarrow$ {\sc Assemble}$\left( v_r ;~~n^{\textrm{\tiny{H}}}_{ij} (\gamma_{ij} + {^n}\gamma^{\textrm{\tiny{NL}}}_{ij})_+^{3/2} \tfrac{\mathbf{x}_i - \mathbf{x}_j}{\|\mathbf{x}_i - \mathbf{x}_j\|} (\delta_{ir} -\delta_{jr})  \right)$ for all $v_r\in E_0(X)$
\STATE $\mathbf{K} \leftarrow$ {\sc Assemble}$\left( v_r, v_s ;~~\tfrac{n^{\textrm{\tiny{H}}}_{ij}  (\gamma_{ij} + {^n}\gamma^{\textrm{\tiny{NL}}}_{ij})_+^{3/2} }{\|\mathbf{x}_i - \mathbf{x}_j\|} \mathbf{I} (\delta_{ir} -\delta_{jr})  (\delta_{is}-\delta_{js})~... \right.$  
\\ \hspace{.4in} 
$\left.     - \left[ \tfrac{3 n^{\textrm{\tiny{H}}}_{ij} (\gamma_{ij} + {^n}\gamma^{\textrm{\tiny{NL}}}_{ij})_+^{1/2} }{2} + \tfrac{n^{\textrm{\tiny{H}}}_{ij} (\gamma_{ij} + {^n}\gamma^{\textrm{\tiny{NL}}}_{ij})_+^{3/2}}{ \|\mathbf{x}_i - \mathbf{x}_j\| } \right]   \tfrac{(\mathbf{x}_i - \mathbf{x}_j)\otimes(\mathbf{x}_i - \mathbf{x}_j)}{\|\mathbf{x}_i - \mathbf{x}_j\|^2}  (\delta_{ir} -\delta_{jr}) (\delta_{is}-\delta_{js}) \right)$
\\ \hspace{.4in} for all $v_r\in E_0(X)$ and $v_s\in E_0(X)$
\ENDFOR
\RETURN $\{\mathbf{F} , \mathbf{K}, {^n}\qvec{\gamma}^{\textrm{\tiny{NL}}} \}$
\end{algorithmic}
\end{algorithm}

\section{Post-jamming behavior of compressible granular systems}
\label{Section-PostJamming}

We next study the post-jamming behavior of two granular systems that only differ in the level of compressibility of the particles (i.e., in $\nu$). The geometric configuration of the initial granular bed is the same for both systems, and simulations are performed for both Hertz theory and the nonlocal contact formulation (each simulation takes several months of CPU time). Deformations at the particle scale are within the validated range of the nonlocal contact formulation, that is smaller than 40$\%$ (see Figure~\ref{Fig-1D-Experiment}), for relative densities smaller than 0.90. We investigate the evolution of statistical features of the mechanical coordination number (number of non-zero contact forces between a particle and its neighbors), punch and die-wall pressures, the network of contact forces, and the network of contact radiuses. Error bounds in tables and plots represent the 95\% confidence envelope for fitted coefficients.

%%%%%%%%%%%%%%%%%%%%%%%%%%%%
\subsection{Mechanical coordination number} 
\label{Subsection-Coordination}

\begin{figure}[htbp]
    \centering{
        \includegraphics[scale=0.60]{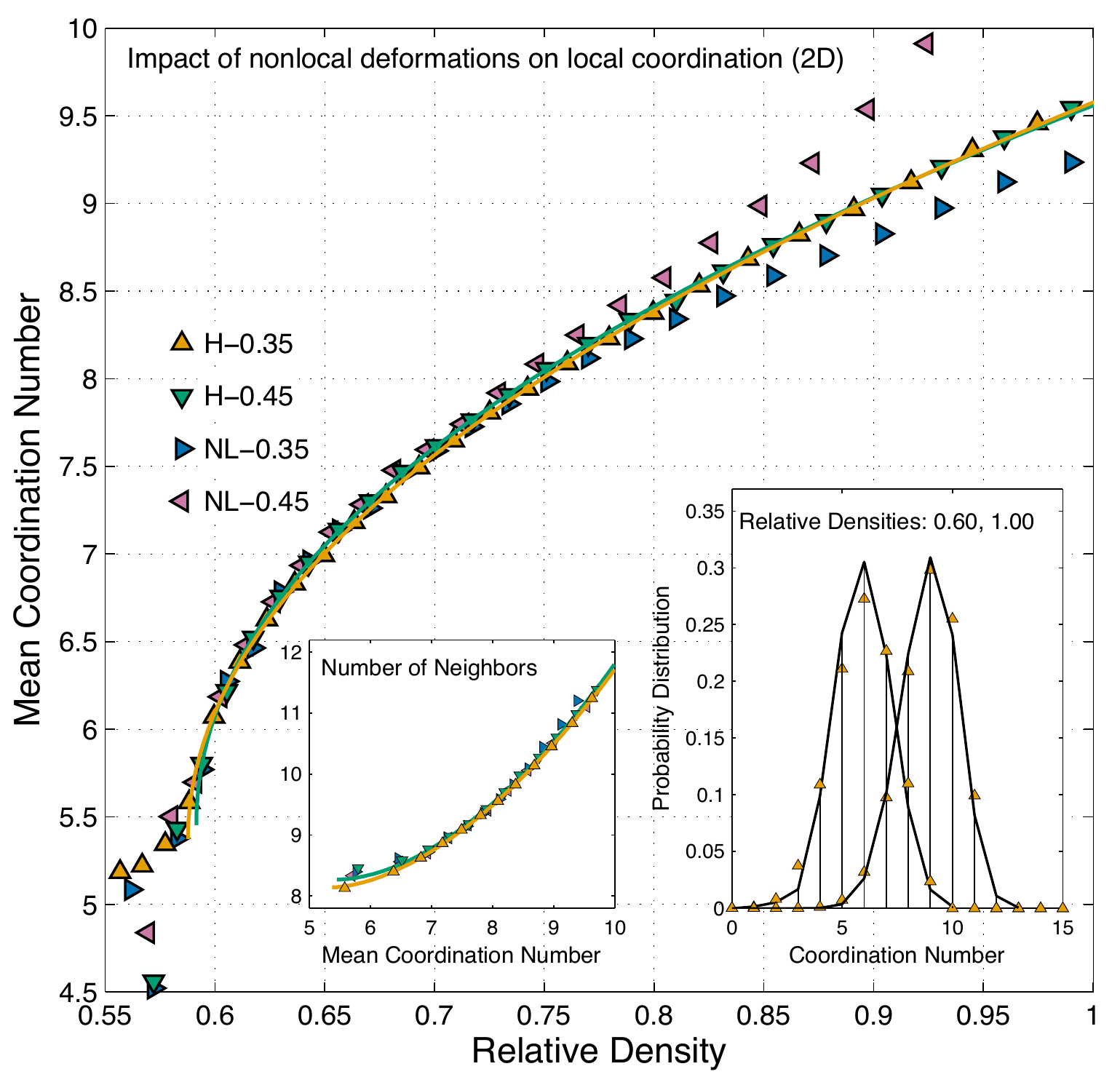}
        \Put(-241,431){\includegraphics[scale=0.28]{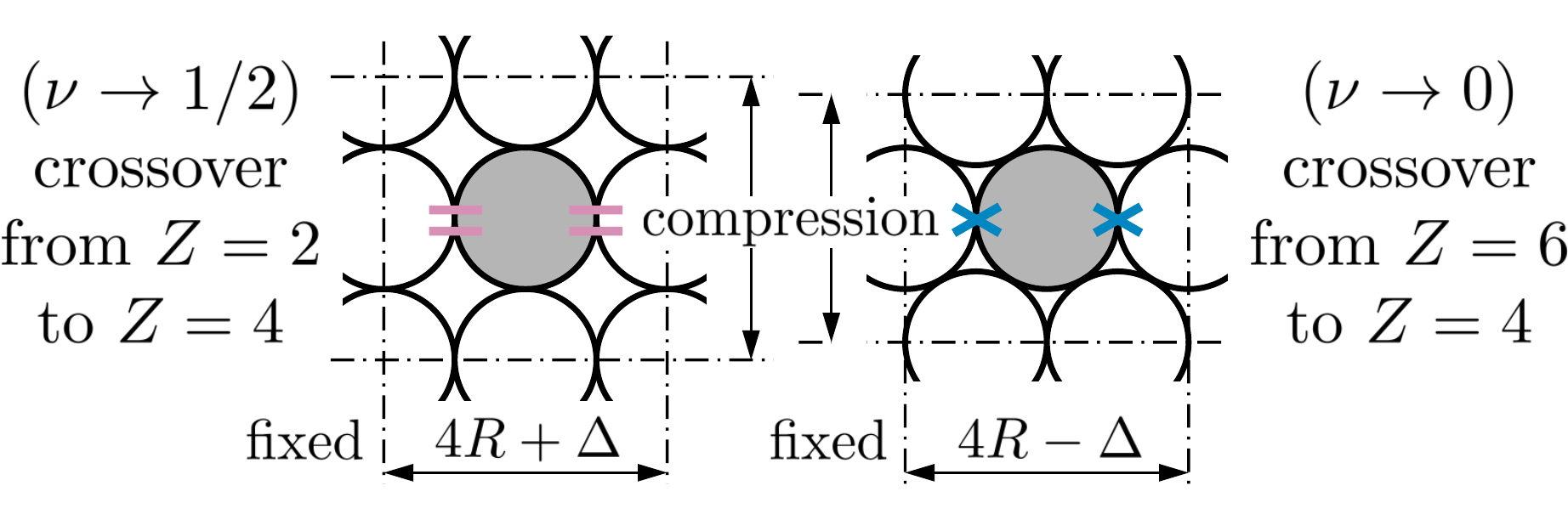}}
    }
    \vspace{-.18in}
\\
\small{
    \begin{tabular}{p{\linewidth}}
    (a)
    \end{tabular}
}
\vspace{.18in}
\\
%\resizebox{0.92\linewidth}{!}
\small{
\begin{tabular}{c|c|ccc|ccc}
    System       & $\rho_c$
                 & $\bar{Z}_c$
                 & $\bar{Z}_0$
                 & $\theta$
                 & $\bar{N}_c$
                 & $\bar{N}_0$
                 & $\varepsilon$
                 \\
                 \hline
    $\nu=0.35$   & $0.5878$
                 & $5.372$
                 & $6.552$
                 & $0.5009$
                 & $8.144$
                 & $0.2547$
                 & $1.723$
                 \\
                 &
                 & $\pm0.2\%$
                 & $\pm0.3\%$
                 & $\pm0.6\%$
                 & $\pm0.2\%$
                 & $\pm3.1\%$
                 & $\pm1.2\%$
                 \\
                 \hline
    $\nu=0.45$   & $0.5916$
                 & $5.452$
                 & $6.347$
                 & $0.4863$
                 & $8.270$
                 & $0.2334$
                 & $1.795$
                 \\
                 &
                 & $\pm0.3\%$
                 & $\pm0.6\%$
                 & $\pm1.0\%$
                 & $\pm0.2\%$
                 & $\pm4.4\%$
                 & $\pm1.6\%$
\end{tabular}
}
\vspace{-.18in}
\\
\small{
    \begin{tabular}{p{\linewidth}}
    (b)
    \end{tabular}
}
\caption{(Color online) (a) Statistical description of the coordination number as a function of relative density. Solid lines correspond to the best fit of equation (\ref{Eqn-CoordinationFit}) to the mean coordination number obtained from the particle contact mechanics simulation of the granular bed (symbols). Bottom-right insert depicts local coordination number probability distribution, and its binomial function approximation. Bottom-left insert depicts the evolution of number of closest neighbors with mean coordination number. Top insert shows two-dimensional examples of coordination number crossover, due to nonlocal deformations, for different levels of compressibility and confinement. (b) Coefficients and critical exponents for best fits.}
\label{Fig-CN-Cylinder}
\end{figure}

The mean coordination number $\bar{Z}$ evolves as a power law of the following form
\begin{equation}
    \bar{Z} - \bar{Z}_c
    =
    \bar{Z}_0 (\rho-\rho_c)^{\theta}
%    +
%    \bar{Z}_{\textrm{\tiny{NL}}} \left[\tanh(\tfrac{\rho-\rho_m}{k_{\textrm{\tiny{NL}}}})+1\right]
    \label{Eqn-CoordinationFit}
\end{equation}
where $\rho_c$ is the critical relative density, $\bar{Z}_c$ is the minimal average coordination number and $\theta$ is the critical exponent. This well-known critical-like behavior has an exponent consistent with $1/2$ for different pair-interaction contact laws, polydispercity and dimensionality of the problem \cite{OHern-2002,OHern-2003,Durian-1995}. Figure~\ref{Fig-CN-Cylinder}(a) shows the results obtained from the particle contact mechanics simulations for Hertz theory and their best fit to equation (\ref{Eqn-CoordinationFit}). For $\nu=0.35$, jamming occurs at $\bar{Z}_c=5.372$ and  $\rho_c=0.5878$ with $\theta=0.5009$. The fit to numerical results is excellent not only near jamming but also at large relative densities [see Figure~\ref{Fig-CN-Cylinder}(b)]. It is worth noting that the isostatic condition for frictionless packings implies a critical coordination number equal to 6 and a critical density close to 0.64. Our initial powder bed, however, is generated such that a subset of particles is jammed, namely the backbone, while the remainder of the particles are not jammed but locally imprisoned by their neighbors, referred to as rattlers. In this work, we do not exclude rattlers from packings, when calculating packing densities, mean coordination numbers and networks of contact forces. Therefore, compaction is initially characterized by rearrangement and very small deformations of the particles, and it is followed by an early jamming transition at $\bar{Z}_c<6$, reaching the isostatic coordination number at a relative density very close to 0.60. There exists a body of work that indicates that $\rho_c$ depends on the protocol used for obtaining jammed configurations and that monodisperse systems are prone to crystallization \cite{Baranau-2104, Chaudhuri-2010, Schreck-2011, Vagberg-2011}. While a detailed analysis concerning these points is a worthwhile direction of future work, specially the investigation of polydisperse packings for which some results are observers to be independent of the generation protocol, here we restrict our discussion to post-jamming behavior and to one preparation protocol. Results for $\nu=0.45$ are only marginally different.

Hertz theory, and any other contact law formulated locally as pair-interactions, does not account for the interplay of deformations due to multiple contact forces acting on a single particle. As relative density increases, these nonlocal deformations are responsible for the formation and breakage of contacts between particles which in turn manifest as a crossover in the local coordination number [see top insert in Figure~\ref{Fig-CN-Cylinder}(a) for two bi-dimensional examples]. This local crossover is predicted by the nonlocal contact formulation and it emerges in the evolution of the mean mechanical coordination number depicted in Figure~\ref{Fig-CN-Cylinder}(a). For highly compressible materials ($\nu=0.35$), the evolution of $\bar{Z}$ is below the predictions of Hertz theory, i.e., some contacts are broken as the level of confinement increases. In sharp contrast, for nearly incompressible materials ($\nu=0.45$), the evolution of $\bar{Z}$ is above the predictions of Hertz theory, i.e., new contacts are formed due to volume conservation at the particle scale. Therefore, while equation (\ref{Eqn-CoordinationFit}) is valid and independent of any material property near jamming, its validity fails at high levels of confinement where $\bar{Z}$ strongly depends on $\nu$.

The probability distribution of the local coordination number $Z$ is best described by the granocentric model proposed in \cite{Clusel-2009,Newhall-2011}. Here we restrict our attention to monodisperse systems and approximate $Z$ by a binomial distributed variable characterized by its mean and variance, that is
\begin{equation}
    Z
    \sim
    B(\lceil\bar{N}\rceil,p)
    =
    B(\lceil\bar{Z}/(1-\textrm{Var}(Z)/\bar{Z})\rceil,1-\textrm{Var}(Z)/\bar{Z})
\end{equation}
where $\bar{N}$ represents the mean number of (closest) neighbors and $p$ corresponds to the probability of finding a contact among these neighbors. Inserts in Figure~\ref{Fig-CN-Cylinder}(a) show that this is an accurate approximation for both formulations at small and large relative densities---though distributions for the nonlocal contact formulation are not shown in the figure. The numerical results also suggest the existence of an universal relationship between $\bar{N}$ and $\bar{Z}$ of the following form
\begin{equation}
    \bar{N}-\bar{N}_c
    =
    \bar{N}_0 (\bar{Z}-\bar{Z}_c)^{\varepsilon}
    \label{Eqn-NeighborsFit}
\end{equation}
with $\bar{N}_c\approx8$ and $\varepsilon\approx7/4$, for both formulations and Poisson's ratios [see Figure~\ref{Fig-CN-Cylinder}(b) for exact numerical values obtained for hertzian interactions].

\begin{figure}[htbp]
\centering{
	\begin{tabular}{ll}
        \includegraphics[scale=0.60]{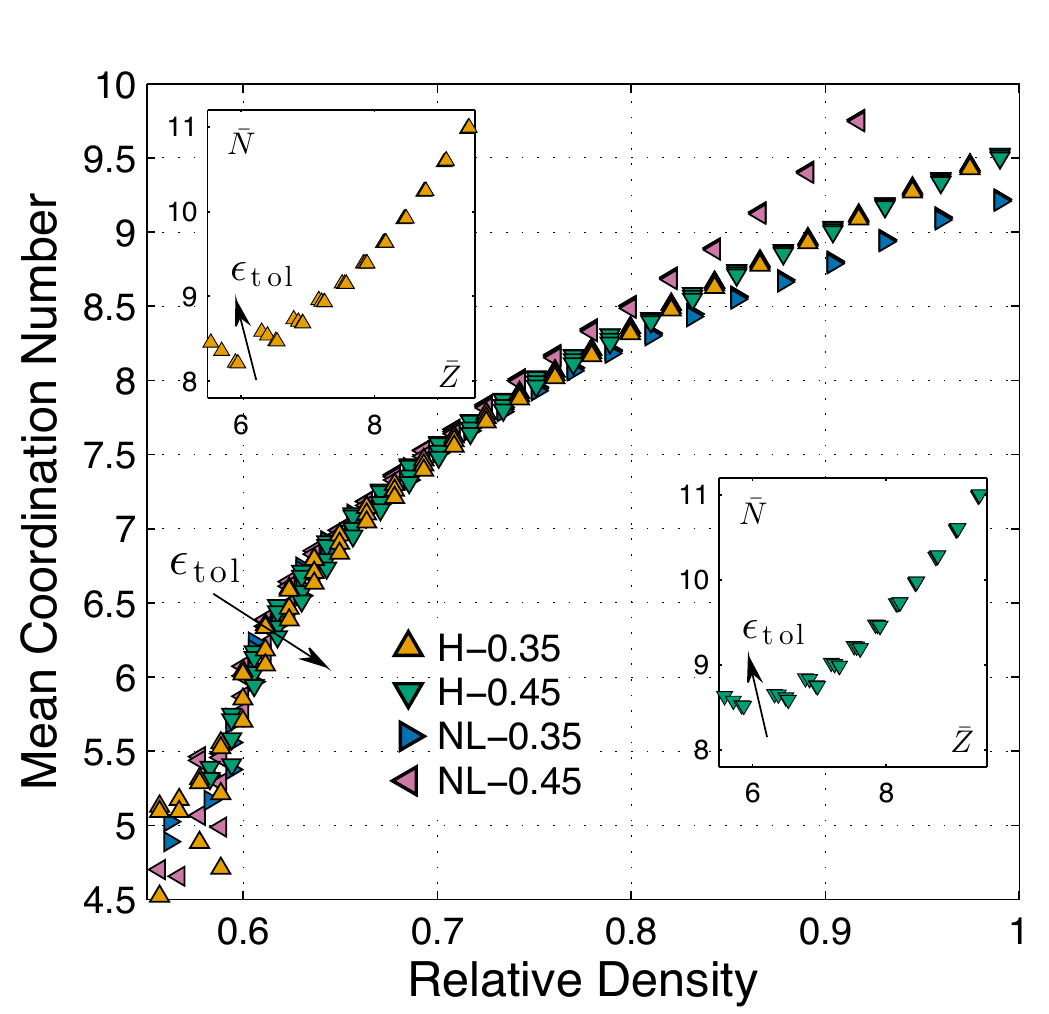}
	&
        \includegraphics[scale=0.60]{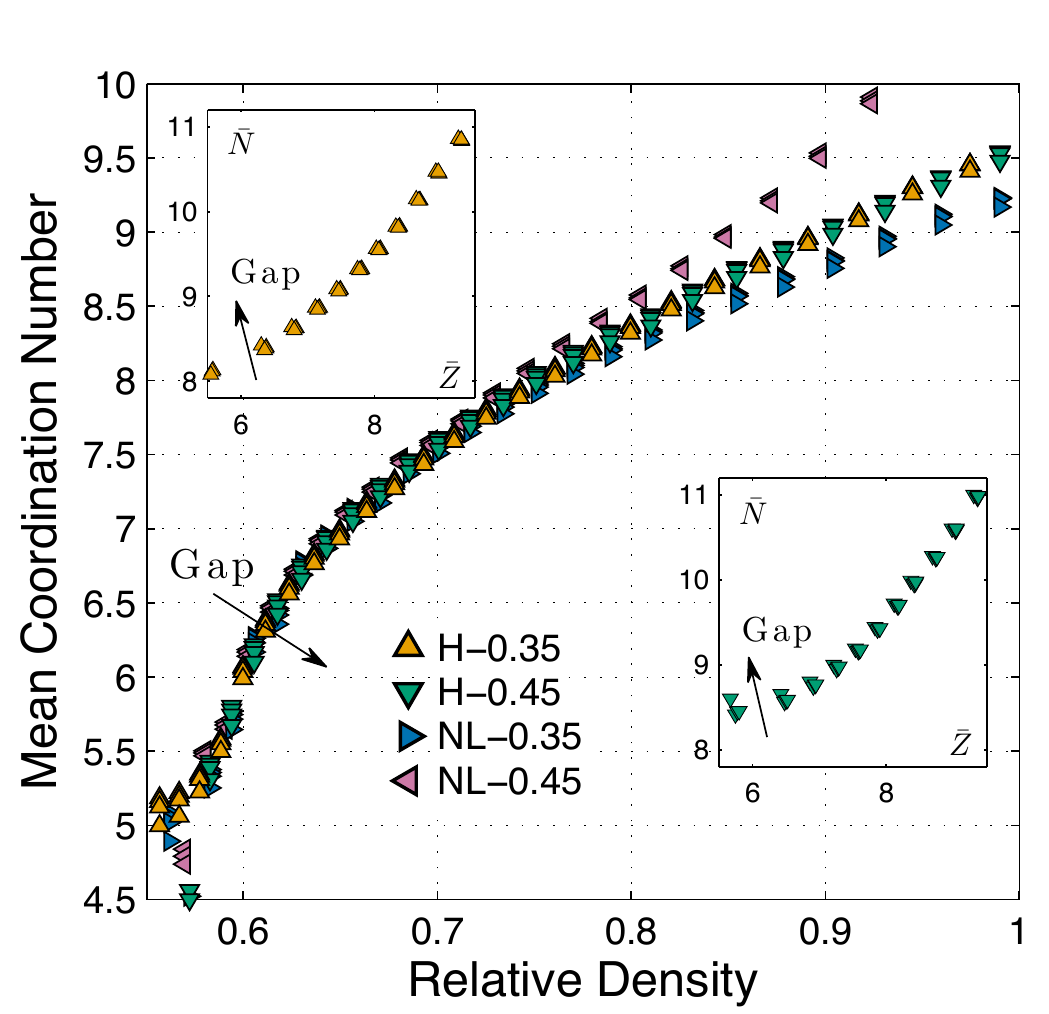}	
    	\vspace{-.18in}
    	\\
	\small{(a)}
   	&
    	\small{(b)}
        \end{tabular}
}
\vspace{-.06in}
\caption{(Color online) (a) Sensitivity of the mean mechanical coordination number to value of the smallest contributing contact force, for $\epsilon_\mathrm{tol}=\gamma/2R \in [0.1\%,0.05\%,0.01\%,0.005\%]$ and $\mathrm{Gap}=4R$. (b) Sensitivity of the mean mechanical coordination number to boundary effects, for $\mathrm{Gap}=[R,2R,3R,4R]$ and $\epsilon_\mathrm{tol}=0.005\%$.}
\label{Fig-CN-Cylinder-Sensitivity}
\end{figure}

Finally, we investigate the sensitivity of these results to the numerical tolerance used to identify the non-zero contact forces that contribute to the mechanical coordination number. We thus define $\epsilon_\mathrm{tol}=(\gamma+ \gamma^{\textrm{\tiny{NL}}})/2R$ as the inter-particle deformation above which a contact force is identified as non-zero. We also investigate how the ordering effect of the container walls penetrates into the bulk of the powder bed. We thus define $\mathrm{Gap}$ as the distance between the boundary and the bulk region. Figure \ref{Fig-CN-Cylinder-Sensitivity}(a) shows that the evolution of the mean coordination number is insensitive to the choice of $\epsilon_\mathrm{tol}$ at high relative densities and barely sensitive at small densities, for values of $\epsilon_\mathrm{tol}$ equal to $0.1\%$, $0.05\%$, $0.01\%$ and $0.005\%$ and for $\mathrm{Gap}=4R$. Figure \ref{Fig-CN-Cylinder-Sensitivity}(b) indicates that the effect of the boundary on the mean coordination number only penetrates one or two particle radiuses into the bulk, for $\epsilon_\mathrm{tol}$ equal to $0.005\%$. Therefore, the results presented in Figure \ref{Fig-CN-Cylinder} correspond to $\epsilon_\mathrm{tol}=0.005\%$ and $\mathrm{Gap}=4R$.

%%%%%%%%%%%%%%%%%%%%%%%%%%%%

\subsection{Applied pressure} 
\label{Subsection-Pressure}

\begin{figure}[htbp]
    \centering{
        \includegraphics[scale=0.64]{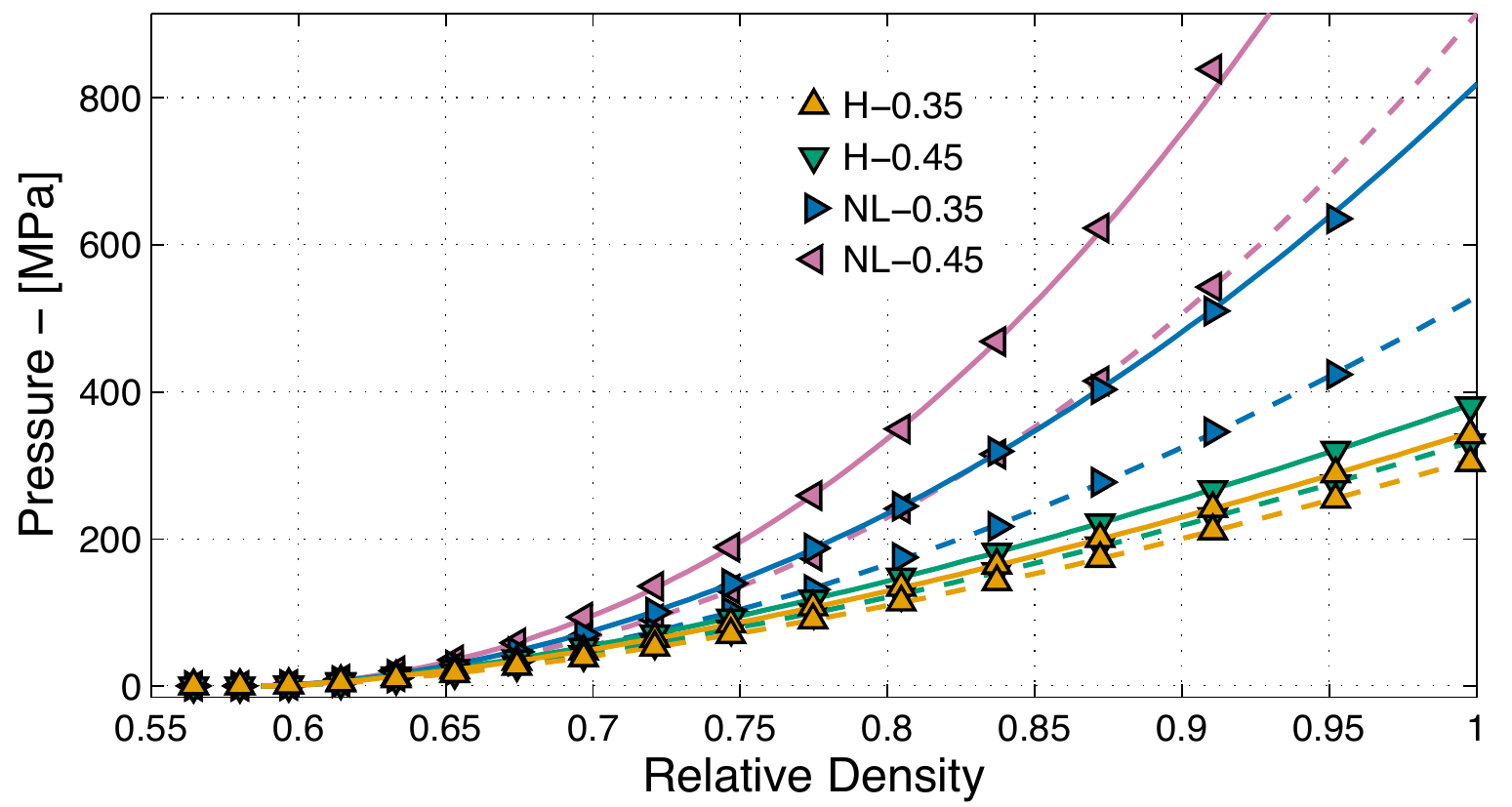}
        \Put(-252,200){\includegraphics[scale=0.55]{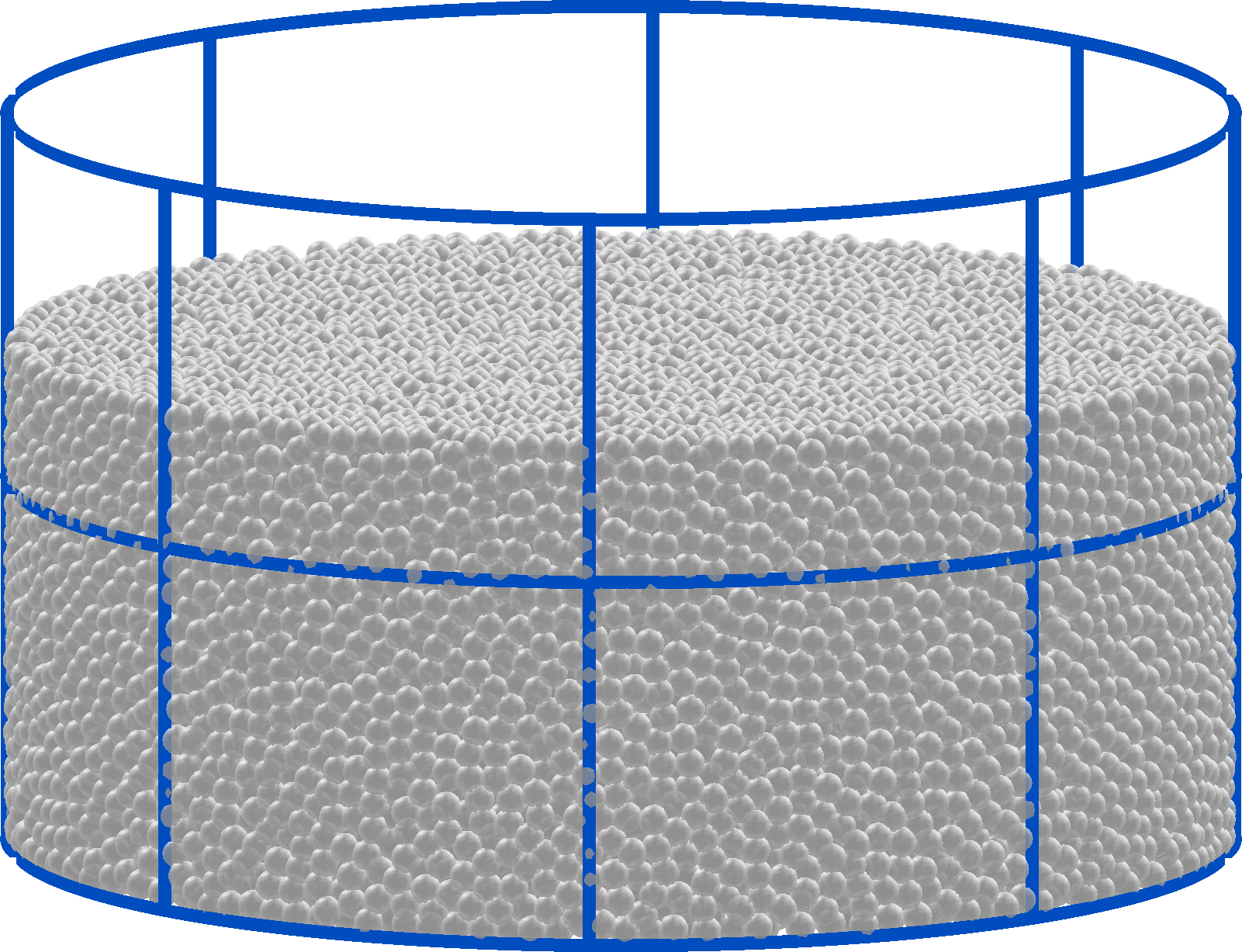}}
    }
\\
\vspace{-.18in}
\small{
    \begin{tabular}{p{\linewidth}}
    (a)
    \end{tabular}
}
\vspace{.18in}
\\
%\resizebox{0.72\linewidth}{!}
\small{
    \begin{tabular}{c|cc|c|cc}
        Pressure     & $K_{\textrm{\tiny{H}}}$
                     & $\beta_{\textrm{\tiny{H}}}$
                     & eqn.(\ref{Eqn-H-Coefficients})
                     & $K_{\textrm{\tiny{NL}}}$
                     & $\beta_{\textrm{\tiny{NL}}}$
                     \\
                     \hline
        Punch        & $1301$
                     & $1.491$
                     & $1269$
                     & $3466$
                     & $2.254$
                     \\
        ($\nu=0.35$) & $\pm0.3\%$
                     & $\pm0.2\%$
                     &
                     & $\pm1.4\%$
                     & $\pm0.5\%$
                     \\
                     \hline
        Wall         & $1237$
                     & $1.565$
                     & ---
                     & $1372$
                     & $2.066$
                     \\
        ($\nu=0.35$) & $\pm0.2\%$
                     & $\pm0.1\%$
                     &
                     & $\pm3.2\%$
                     & $\pm1.2\%$
                     \\
                     \hline
        Punch        & $1454$
                     & $1.498$
                     & $1397$
                     & $8640$
                     & $2.453$
                     \\
        ($\nu=0.45$) & $\pm0.2\%$
                     & $\pm0.1\%$
                     &
                     & $\pm1.5\%$
                     & $\pm0.5\%$
                     \\
                     \hline
        Wall         & $1306$
                     & $1.538$
                     & ---
                     & $5436$
                     & $2.528$
                     \\
        ($\nu=0.45$) & $\pm0.6\%$
                     & $\pm0.3\%$
                     &
                     & $\pm2.1\%$
                     & $\pm0.7\%$
    \end{tabular}
}
\\
\vspace{-.18in}
\small{
    \begin{tabular}{p{\linewidth}}
    (b)
    \end{tabular}
}
\caption{(Color online) (a) Pressures at upper/lower punch (solid lines) and at die wall (dashed lines) as a function of relative density. Lines correspond to the best fit of equation (\ref{Eqn-PressureFit}) to results of the particle contact mechanics simulation of the granular bed (symbols). The insert depicts the compacted granular bed at $\rho=1$. (b) Coefficients and exponents for best fits. Stiffnesses are in MPa.}
\label{Fig-DieForce-Cylinder}
\end{figure}

The pressures applied by the punches and the reaction at the die wall are effective macroscopic variables predicted by the particle contact mechanics simulation. These predictions are presented in Figure~\ref{Fig-DieForce-Cylinder}(a) and follow a power law of the following form
\begin{equation}
    \sigma
    =
    K_{\textrm{\tiny{H}}} (\rho-\rho_c)^{\beta_{\textrm{\tiny{H}}}}
    +
    K_{\textrm{\tiny{NL}}} (\rho-\rho_c)^{\beta_{\textrm{\tiny{NL}}}}    
    \label{Eqn-PressureFit}
\end{equation}
where the pressure $\sigma$ is evaluated at the deformed configuration, $\rho_c$ is obtained from the evolution of $\bar{Z}$, the coefficients in the first term of the equation (i.e., $K_{\textrm{\tiny{H}}}$ and $\beta_{\textrm{\tiny{H}}}$) are best-fitted to the numerical results for Hertz theory, and the remaining coefficients (i.e., $K_{\textrm{\tiny{NL}}}$ and $\beta_{\textrm{\tiny{NL}}}$) are best-fitted to predictions of the nonlocal contact formulation [see Figure~\ref{Fig-DieForce-Cylinder}(b) for numerical values]. It is interesting to note that stiffnesses $K_{\textrm{\tiny{H}}}$ and critical exponents $\beta_{\textrm{\tiny{H}}}$ for hertzian systems are well approximated by
\begin{eqnarray}
    K_{\textrm{\tiny{H}}}~=&~\frac{E}{2\pi(1-\nu^2)}      \textrm{\hspace{.10in};\hspace{.39in}} \beta_{\textrm{\tiny{H}}}~=&~3/2
    \label{Eqn-H-Coefficients}
    \label{Eqn-NL-Coefficients}
\end{eqnarray}
near jamming (see, e.g., \cite{OHern-2002,Durian-1995,Makse-2000,Mason-1997}) and at large relative densities. Also, $\beta_{\textrm{\tiny{H}}} = 3/2$ and $\beta_{\textrm{\tiny{NL}}} \ge 2$ are consistent with the exponents in a series expansion of the pressure applied in the configuration presented in Figure~\ref{Fig-1D-Experiment} (see equation 12 in \cite{Gonzalez-2012}). An analytical derivation of $K_{\textrm{\tiny{NL}}}$ is a worthwhile direction of future work, for which one may expect $K_{\textrm{\tiny{NL}}}/K_{\textrm{\tiny{H}}}$ to only depend on $\nu$ and thus to be in agreement with studies presented in \cite{Gonzalez-2012} for granular crystals or highly packed granular lattices. In a similar manner, one may expect that the behavior observed in Figure~\ref{Fig-CN-Cylinder} for $\bar{Z}$ solely depends on $\nu$. A detailed analysis concerning this point is however beyond the scope of this work.

%%%%%%%%%%%%%%%%%%%%%%%%%%%%
\subsection{Network of contact forces} 
\label{Subsection-NetworkCF}

\begin{figure}[htbp]
\centering{
\begin{tabular}{ll}
    \includegraphics[scale=0.64]{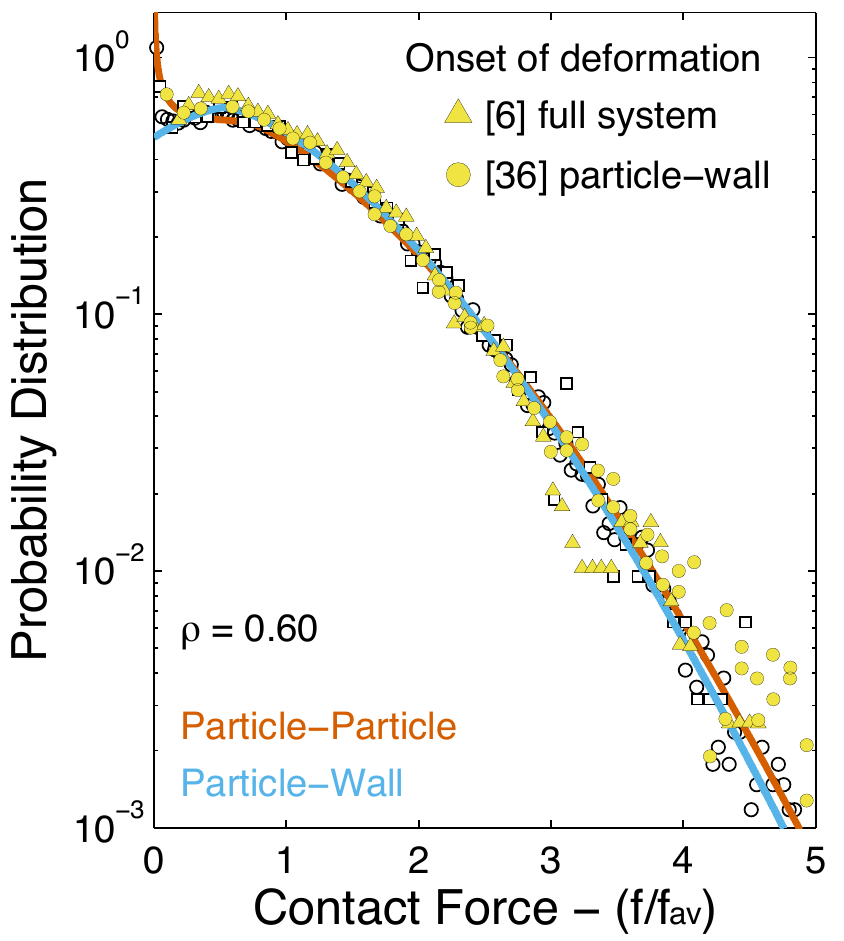}
    &
    \includegraphics[scale=0.64]{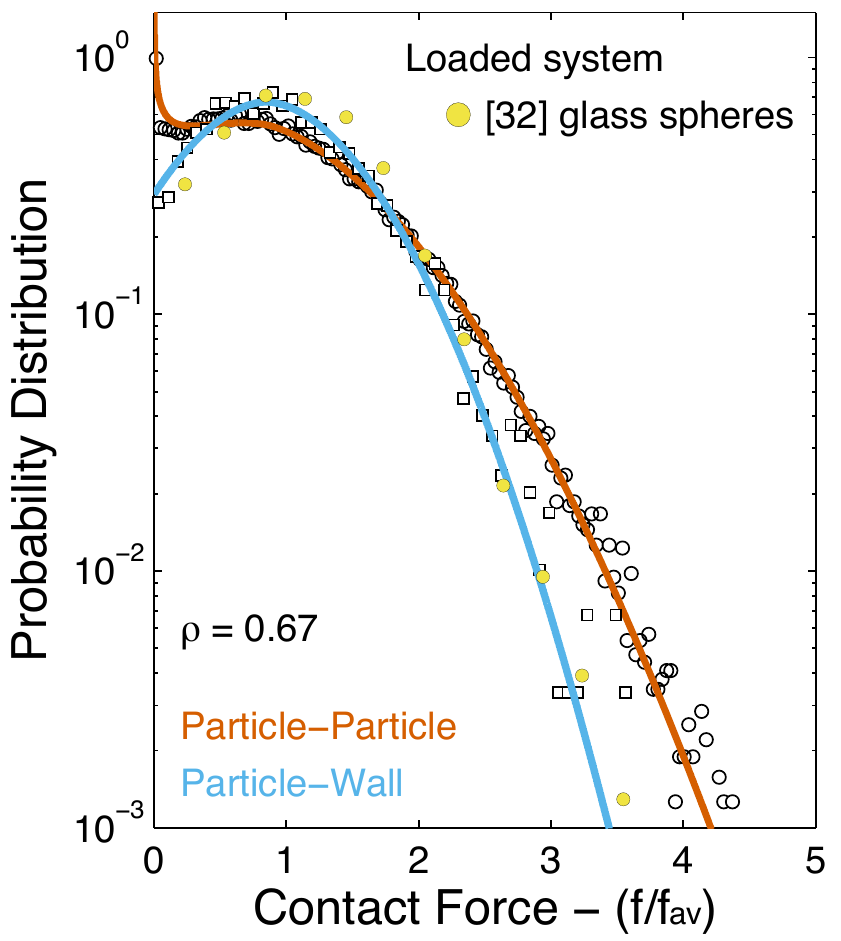}
    \vspace{-.18in}
    \\
    \small{(a)}
    &
    \small{(b)}
    \\
    \includegraphics[scale=0.64]{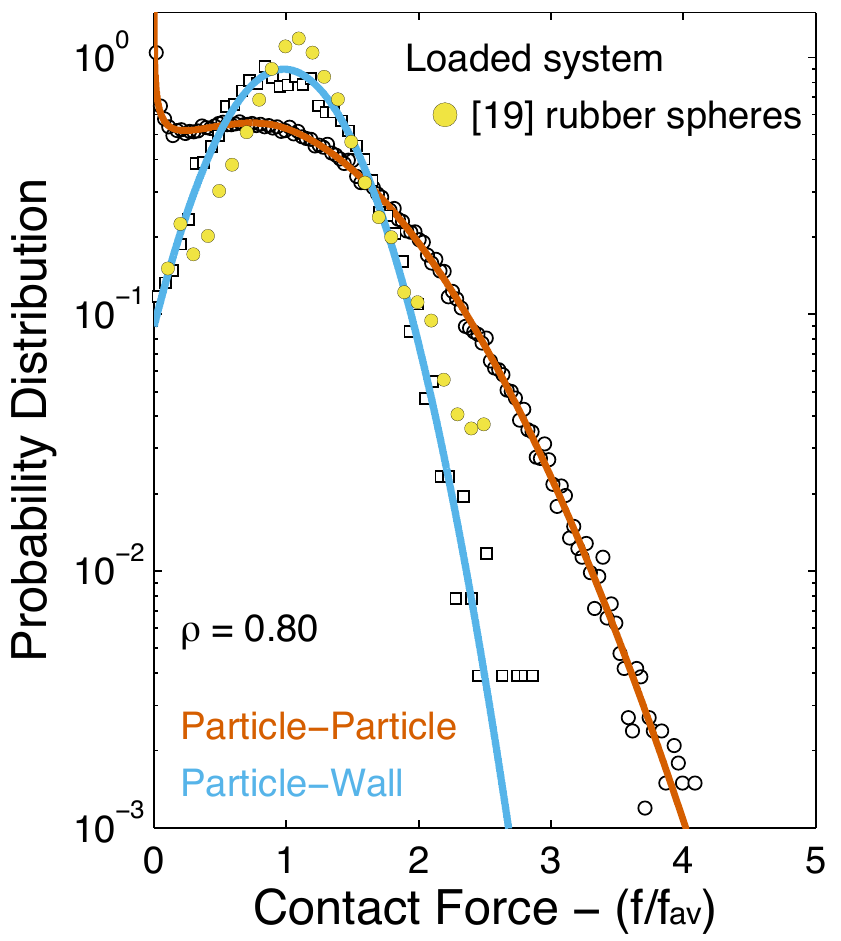}
    &
    \includegraphics[scale=0.64]{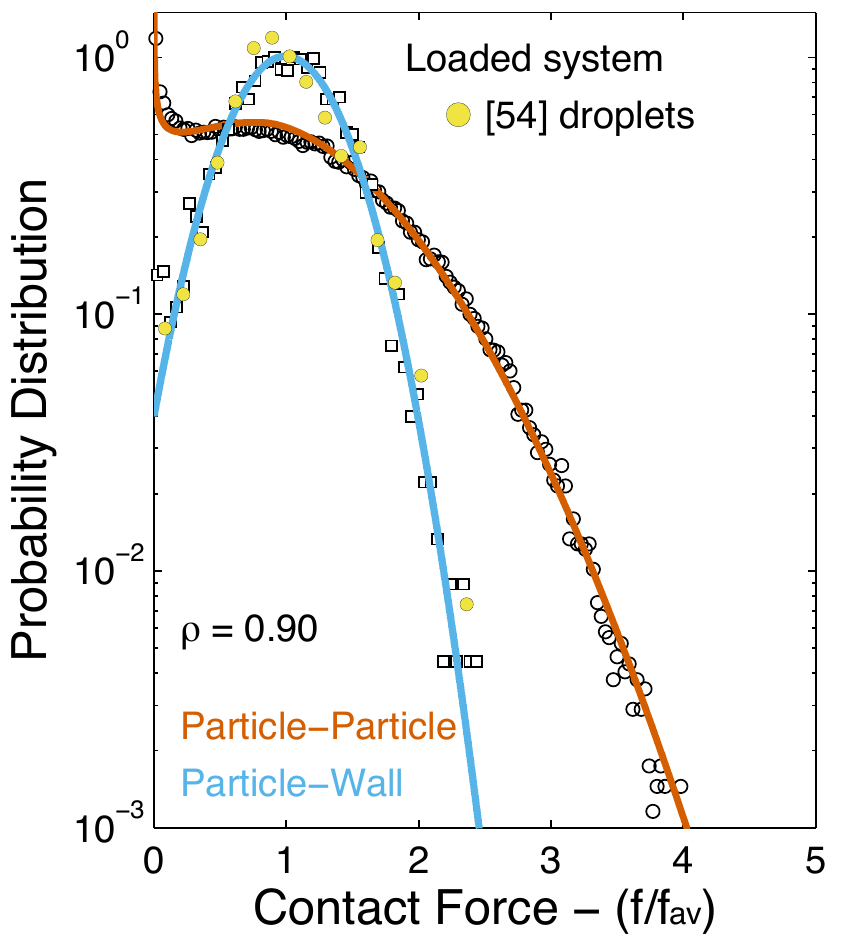}
    \vspace{-.18in}
    \\
    \small{(c)}
    &
    \small{(d)}
\end{tabular}
}
\vspace{-.06in}
\caption{(Color online) Probability distribution of particle-particle ($\circ$) and particle-wall ({\scriptsize $\square$}) contact forces determined from the particle contact mechanics simulation for (a) Hertz theory with $\nu=0.45$, (b) nonlocal contact formulation with $\nu=0.35$, and (c)-(d) nonlocal contact formulation with $\nu=0.45$. Filled symbols correspond to die compaction experiments in cylindrical containers and solid lines correspond to the best fit of the numerical results to equation (\ref{Eqn-ForceFit}), with (\ref{Eqn-PP-Coefficients}) and (\ref{Eqn-PW-Coefficients}). All calculations correspond to $\epsilon_\mathrm{tol}=\gamma/2R=0.005\%$ and particle-particle calculations are for $\mathrm{Gap}=4R$.
%{\color{red} (a) \cite{Brujic-2003a}, \cite{Mueth-1998}, (b) \cite{Makse-2000}, (c) \cite{Erikson-2002}, (d) \cite{Zhou-2006}}
}
\label{Fig-CF-Cylinder}
\end{figure}

The probability distribution of the contact forces is assumed to have an algebraic tail with exponent $\alpha$ and to be translated by $f_0$, that is
\begin{equation}
    P(\bar{f})
    \propto
    \bar{f}^{\gamma-1}
    \exp[-\lambda^{\alpha} | \bar{f} - f_0 |^{\alpha}]
    \label{Eqn-ForceFit}
\end{equation}
where $\bar{f}=f/f_\mathrm{\small av}$ is the non-dimensional contact force and $\gamma$ is a shape parameter that controls the behavior of weak forces---for $\alpha>1$, $\int_0^{+\infty}P(\bar{f})\textrm{d}\bar{f}$ is finite if $\gamma>0$; and for $\gamma=1$, $P(\bar{f}=0)$ is finite.  Contact forces between particles and between particles and walls are studied separately and non-dimensionalized by their own average value $f_\mathrm{\small av}$. Figure \ref{Fig-CF-Cylinder} shows that, as relative density increases, the tails of both probability distributions transition from exponential ($\alpha=1$) to Gaussian ($\alpha=2$). However, the distributions, although similar at jamming onset, are very different at full compaction. The particle-wall force distribution becomes increasingly symmetric, that is its mode approaches its mean value. In sharp contrast, the particle-particle force distribution shows a peak at small forces and a local maximum, or a plateau, whose location saturates to $3/4$. These results agree with previous studies that noted a transition from exponential to Gaussian tails \cite{Antony-2000,Brujic-2003a,Brujic-2003b,Erikson-2002,Jorjadze-2013,Makse-2000,Thornton-1997,vanEerd-2007}. They also extend to three-dimensional systems the observation that distributions measured at the walls do not reflect accurately distributions of the bulk in general (see, e.g., \cite{Snoeijer-2004,Tighe-2008} for two-dimensional studies).

A best fit of the numerical results suggests the following further simplification
\begin{eqnarray}
    \textrm{particle-particle:\hspace{.02in}}~f_0=&\sqrt{\alpha-1} \textrm{\hspace{.05in};\hspace{.15in}} &\lambda=\sqrt{\alpha}-1/2
    \label{Eqn-PP-Coefficients}
    \\
    \textrm{particle-wall:\hspace{.02in}}~f_0=&\alpha-1            \textrm{\hspace{.07in};\hspace{.07in}} &\gamma=1
    \label{Eqn-PW-Coefficients}
\end{eqnarray}
Thus, after best fitting the numerical results of the particle contact mechanics simulations to (\ref{Eqn-ForceFit}) with (\ref{Eqn-PP-Coefficients}-\ref{Eqn-PW-Coefficients}), it is evident from Figure~\ref{Fig-CF-Cylinder} that two parameters fully-describe each probability distribution with remarkable accuracy. Furthermore, the numerical results are in excellent agreement with experimental measurements reported in the literature for one-dimensional compaction of spherical deformable particles in cylindrical containers. These measurements are obtained at, or near, the wall using a carbon paper technique \cite{Erikson-2002,Mueth-1998,Makse-2000}, or confocal microscopy \cite{Zhou-2006}, respectively---with the remarkable exception of \cite{Brujic-2003a}, and recently \cite{Brodu-2015}, which are obtained within the bulk of monodisperse emulsions by confocal microscopy. Naturally, the statistical significance obtained for particle-wall interactions is smaller than the one obtained for particle-particle interactions---e.g., in this study, the ratio between the number of each type of interaction remains almost constant during compaction and equal to 3\%.

\begin{figure}[htbp]
\centering{
    \includegraphics[scale=0.66]{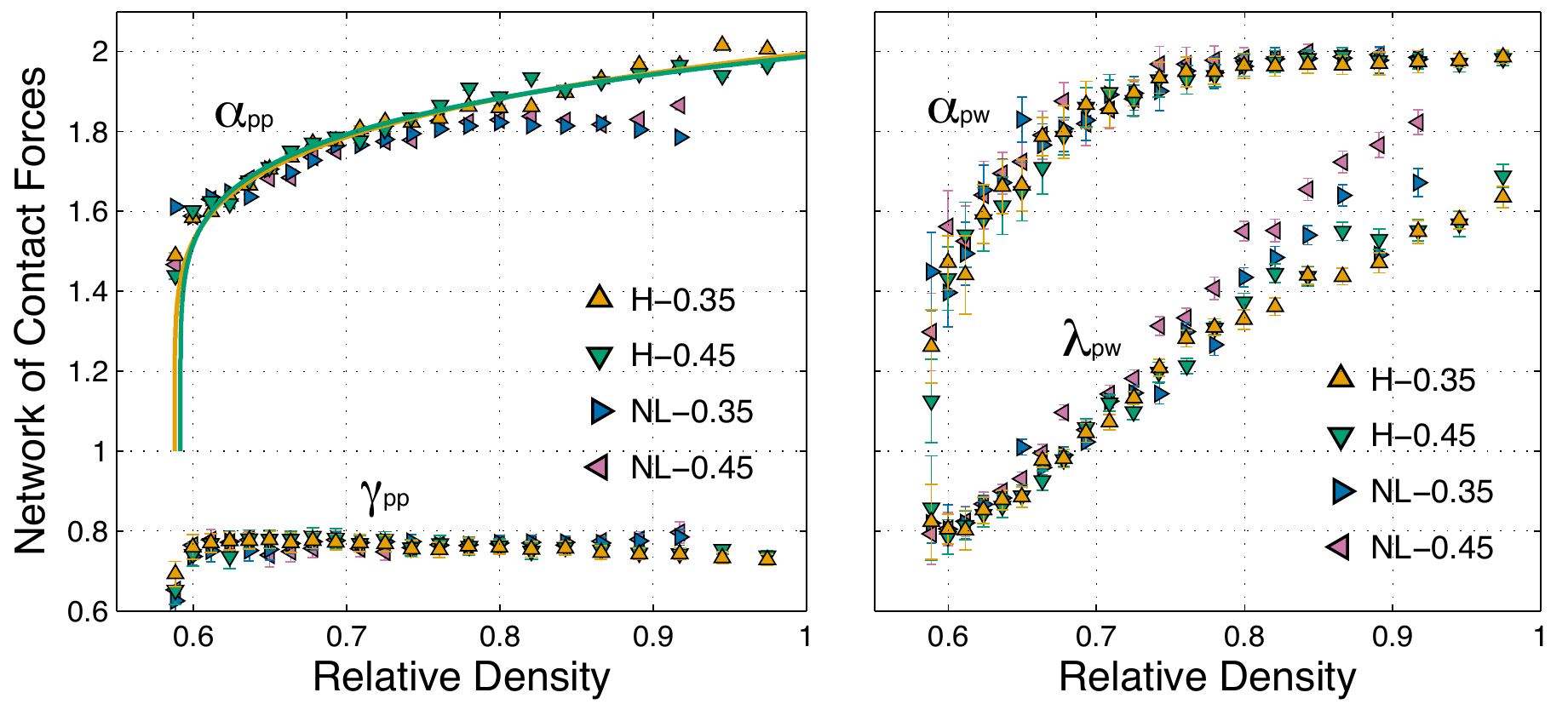} 
}
\caption{(Color online) Statistical description of the probability distribution of particle-particle (PP) and particle-wall (PW) contact forces, as a function of relative density. Symbols represent the parameters obtained from the best fit of numerical results to equation (\ref{Eqn-ForceFit}), with (\ref{Eqn-PP-Coefficients}) and (\ref{Eqn-PW-Coefficients}). All calculations correspond to $\epsilon_\mathrm{tol}=\gamma/2R=0.005\%$ and particle-particle calculations are for $\mathrm{Gap}=4R$.}
\label{Fig-ForceFitEvolution}
\end{figure}

Figure~\ref{Fig-ForceFitEvolution} shows the evolution of $\{\alpha,\gamma\}_{\textrm{\tiny{PP}}}$ for particle-particle interactions and of $\{\alpha,\lambda\}_{\textrm{\tiny{PW}}}$ for particle-wall interactions. Interestingly, the figure suggests a critical-like behavior for the parameter $\alpha_{\textrm{\tiny{PP}}}$ with critical exponent $1/5$ and critical density $\rho_c$. It also reveals that predictions of the nonlocal contact formulation depart from those obtained with Hertz theory at moderate to high levels of confinements (e.g., $\alpha_{\textrm{\tiny{PP}}}$ saturates at 1.8, not reaching truly Gaussian values). The systematic investigation of this critical-like behavior, and the elucidation of the material properties and microstructural features that control it, are worthwhile directions of future research. In addition, the figure revelas that $\gamma_{\textrm{\tiny{PP}}}$ is larger than 0.6 and saturates  at 0.80, whereas $\gamma_{\textrm{\tiny{PW}}}$ is found to be 1. This is in agreement with recent theoretical and numerical calculations (see, e.g., \cite{Bo-2014,Lerner-2013} and references therein). It is worth noting that the probability distribution of particle-wall contact forces thus takes a finite value at zero proportional to $\exp[-\lambda^{\alpha} (\alpha-1)^{\alpha}]$. Finite values for $P(\bar{f}=0)$ have been reported in \cite{Bo-2014} for particle-particle forces.

\begin{figure}[htbp]
\centering{
\begin{tabular}{ll}
    \includegraphics[scale=0.27, clip, trim=375 100 375 100]{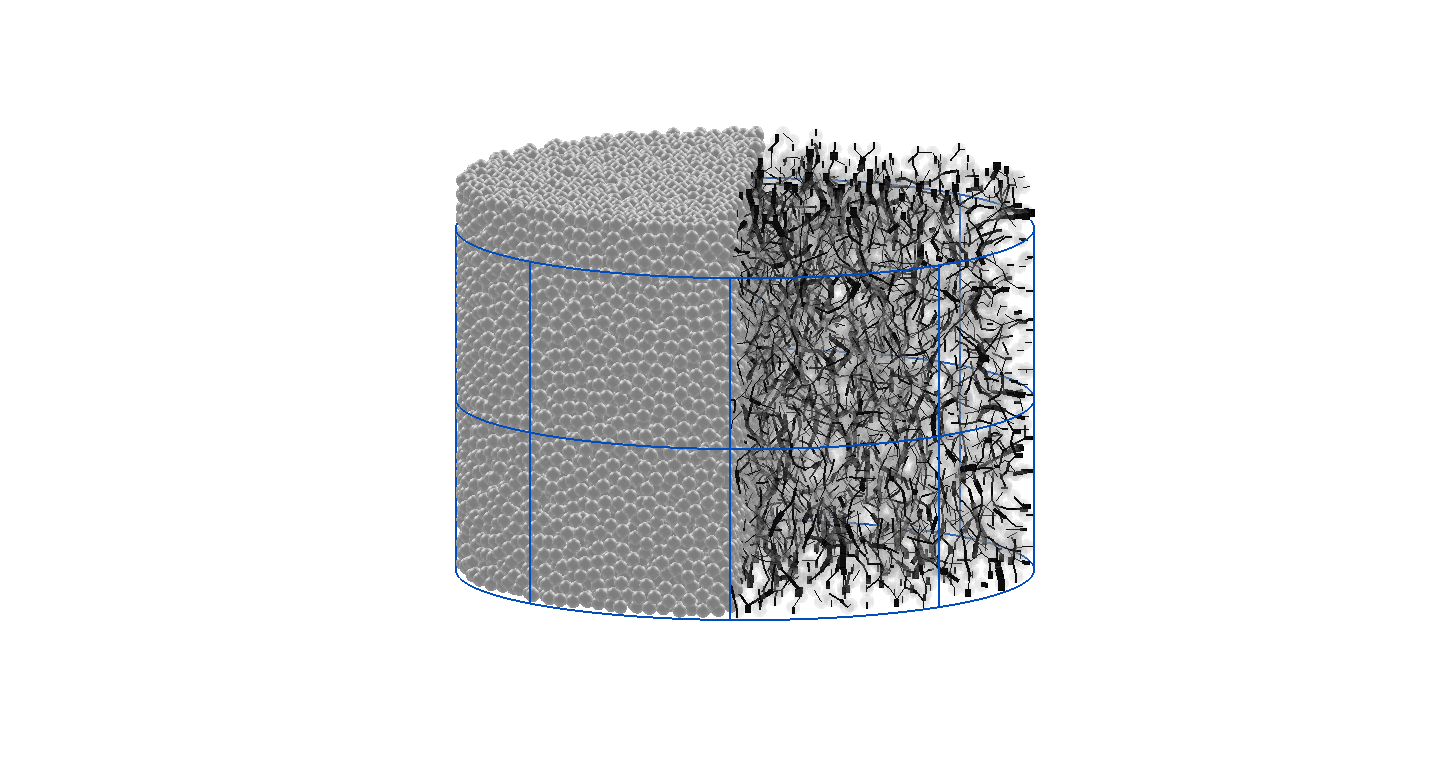}
    &
    \includegraphics[scale=0.27, clip, trim=375 100 375 100]{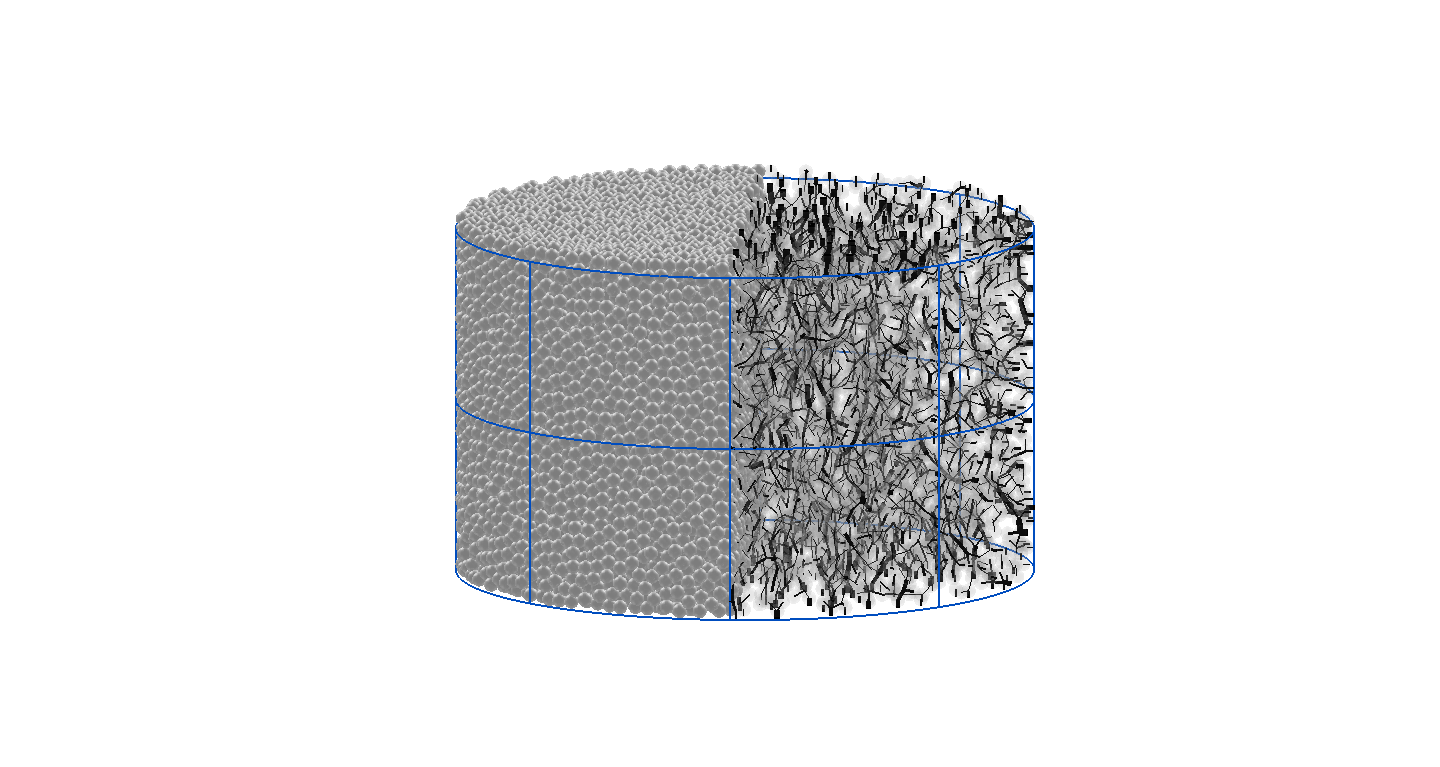}
    \vspace{-.18in}
    \\
    \small{(a)}
    &
    \small{(b)}
    \\
    \includegraphics[scale=0.27, clip, trim=375 100 375 100]{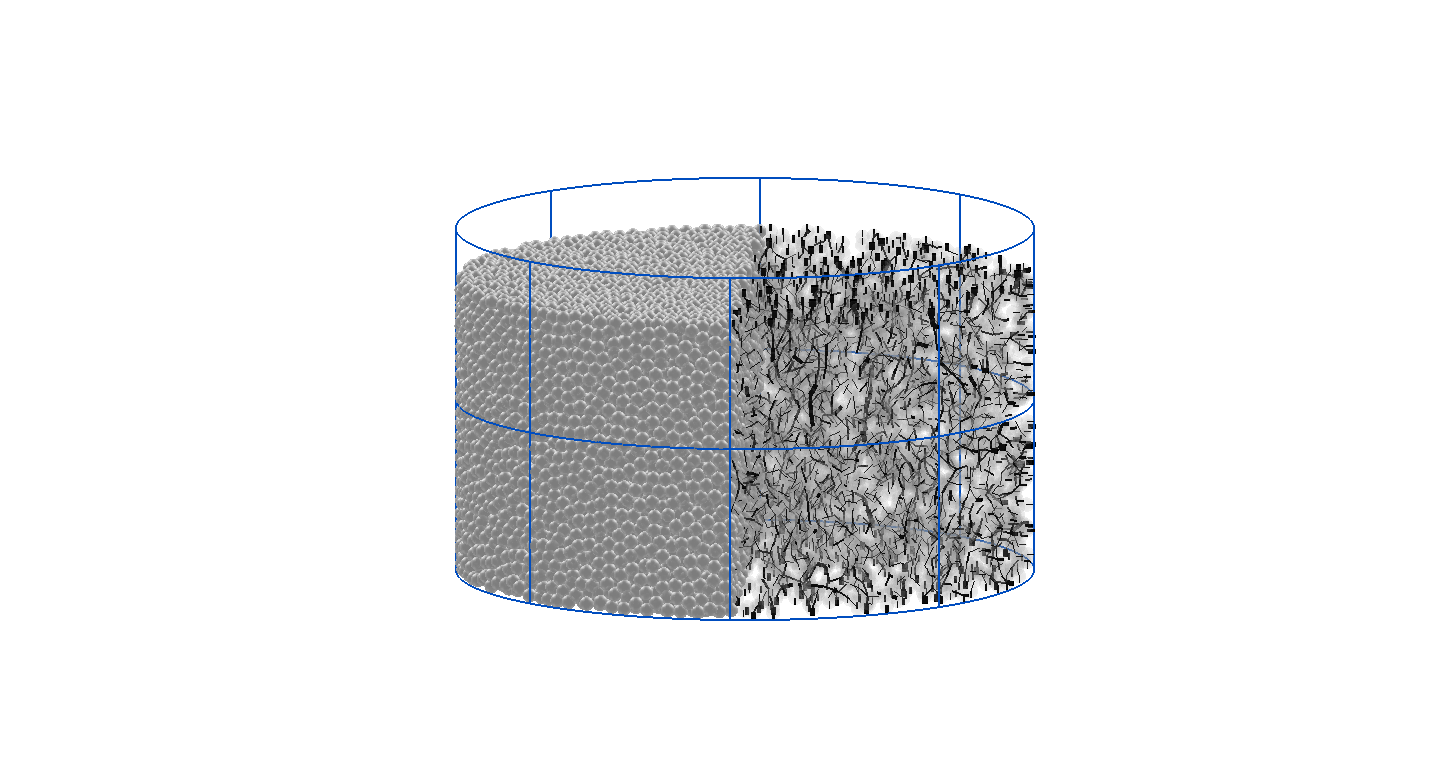}
    &
    \includegraphics[scale=0.27, clip, trim=375 100 375 100]{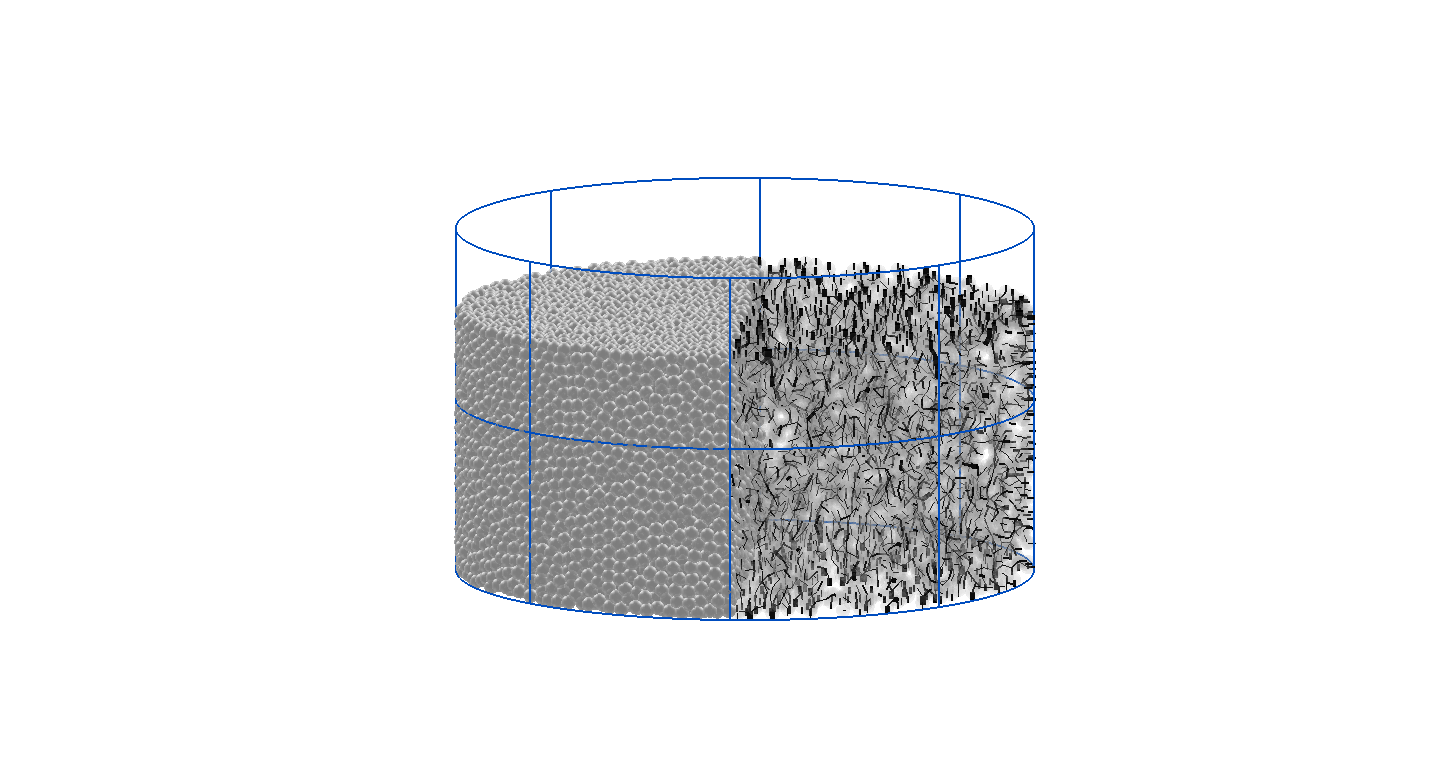}
    \vspace{-.18in}
    \\
    \small{(c)}
    &
    \small{(d)}
\end{tabular}
}
\vspace{-.06in}
\caption{Network of forces in the compacted granular bed for (a) $\rho=0.60$, (b) $\rho=0.67$,  (c) $\rho=0.80$ and (d) $\rho=0.90$. Thicker lines depict larger forces. Only contact forces $\bar{f}$ larger than 2.5 are not shown and the particles involved in these pair-interactions are depicted as semi-transparent gray spheres.
}
\label{Fig-CF-3D-Cylinder}
\end{figure}

Finally, Figure \ref{Fig-CF-3D-Cylinder} illustrates the network of forces in the three-dimensional granular system for four different relative densities. We only visualize non-dimensional contact forces $\bar{f}$ larger than 2.5, that is the forces that carry most of the load in the system. It is clearly observed in the figure that at low relative density or confinement the network is inhomogeneous in space and characterize by a broad range of forces. In contrast, at high relative density or confinement the network becomes more homogeneous in space and the range of observed forces is narrower \cite{Makse-2000}. These observations are consisten with the transition from exponential to Gaussian tails in the distribution of particle-particle interactions under increasing relative density.

%%%%%%%%%%%%%%%%%%%%%%%%%%
\subsection{Network of contact radiuses and areas}
\label{Subsection-NetworkCR}

\begin{figure}[htbp]
\centering{
\begin{tabular}{ll}
    \includegraphics[scale=0.64]{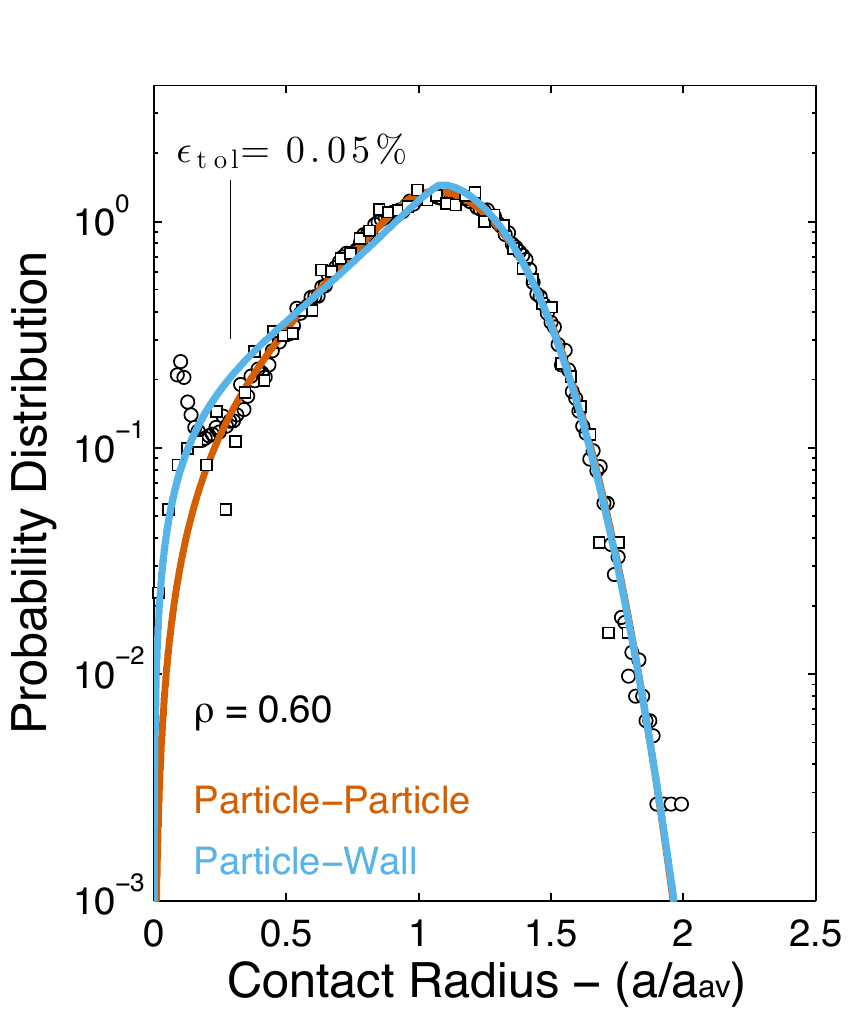}
    &
    \includegraphics[scale=0.64]{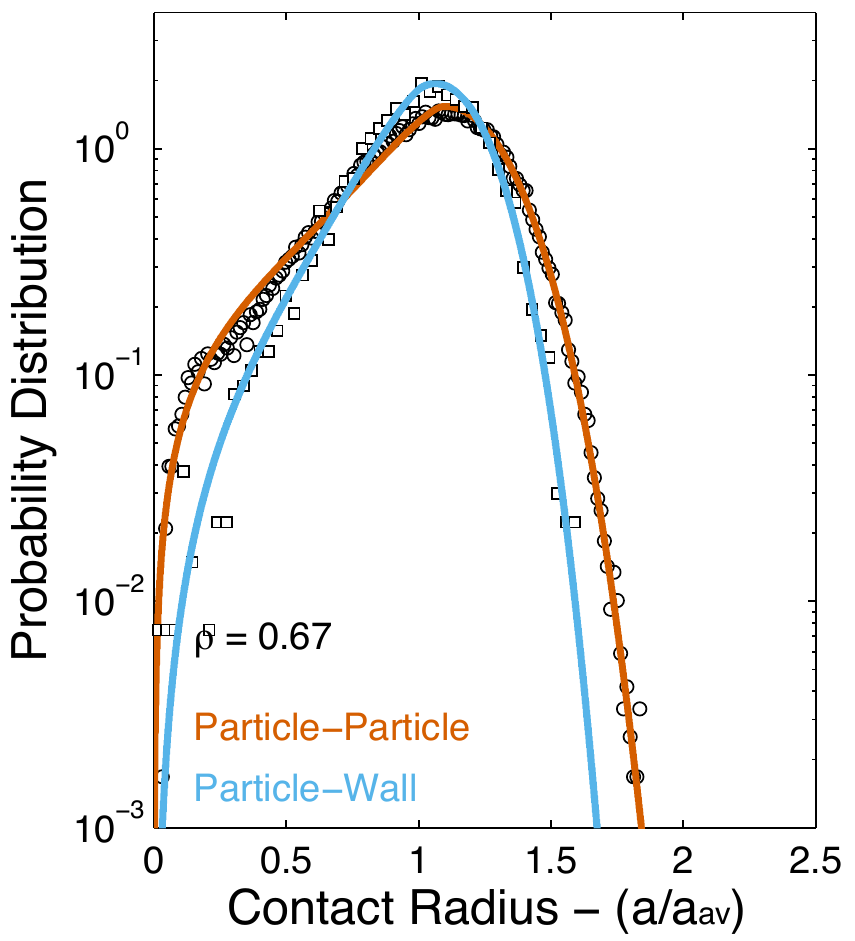}
    \vspace{-.18in}
    \\
    \small{(a)}
    &
    \small{(b)}
    \\
    \includegraphics[scale=0.64]{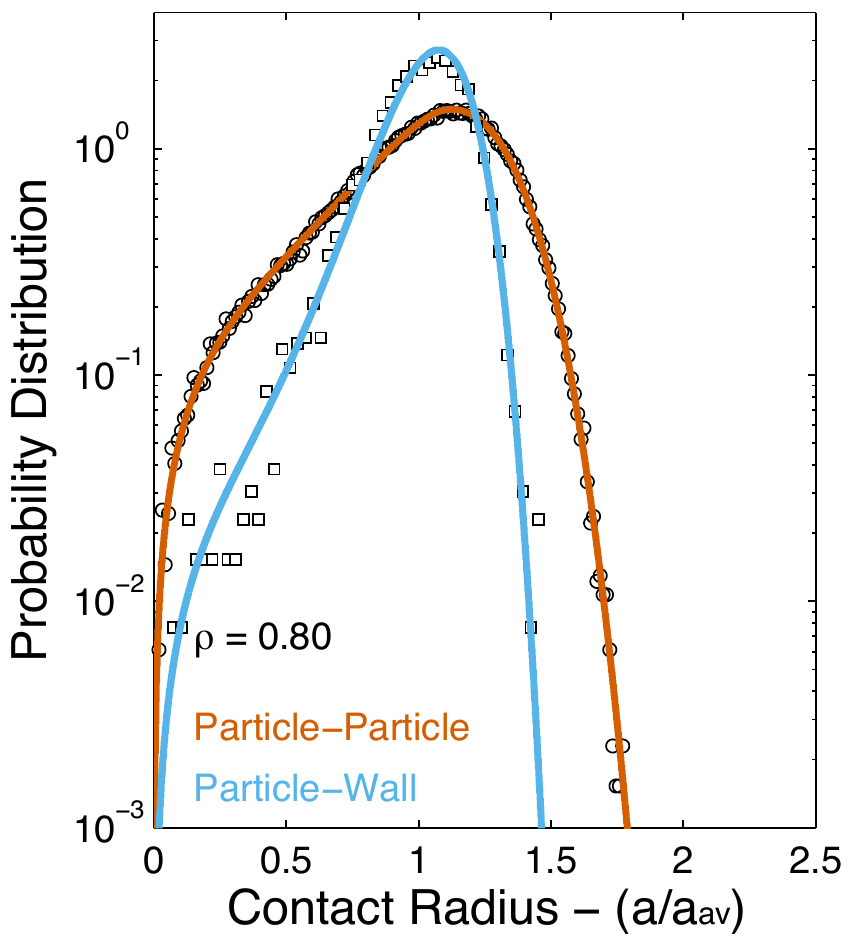}
    &
    \includegraphics[scale=0.64]{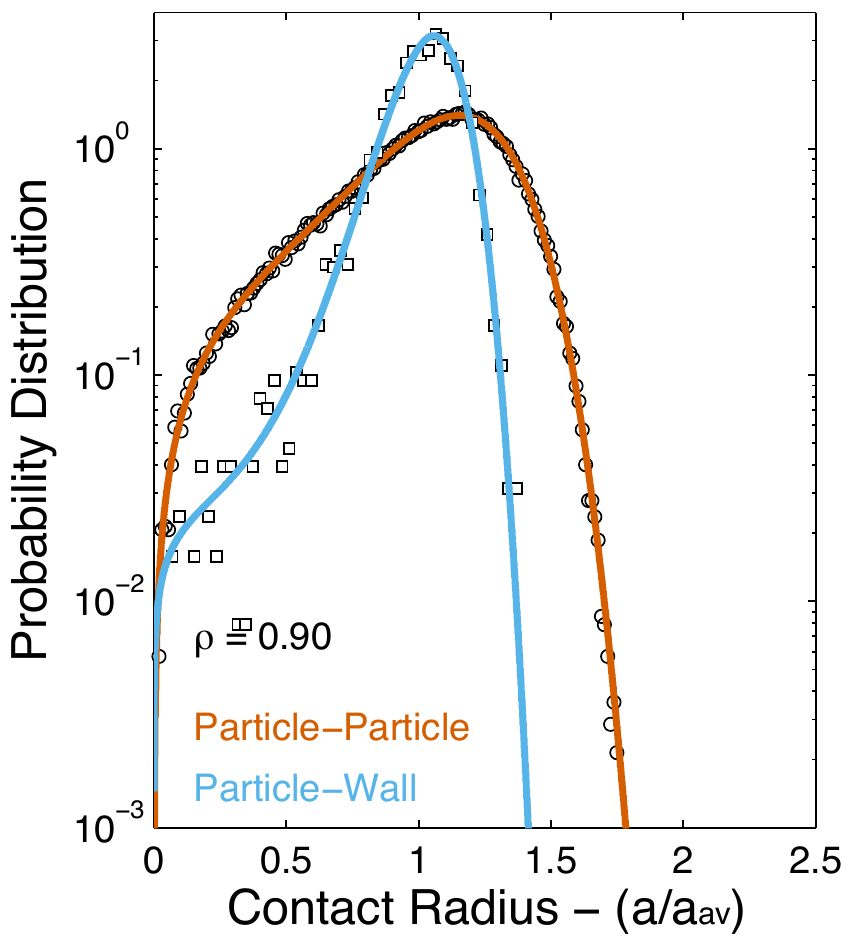}
    \vspace{-.18in}
    \\
    \small{(c)}
    &
    \small{(d)}
\end{tabular}
}
\vspace{-.06in}
\caption{(Color online) Probability distribution of particle-particle ($\circ$) and particle-wall ({\scriptsize $\square$}) contact radiuses determined from the particle contact mechanics simulation for (a) Hertz theory with $\nu=0.45$, (b) nonlocal contact formulation with $\nu=0.35$, and (c)-(d) nonlocal contact formulation with $\nu=0.45$. Solid lines correspond to the best fit of the numerical results to equation (\ref{Eqn-ContactRadiusFit}). All calculations correspond to $\epsilon_\mathrm{tol}=\gamma/2R=0.005\%$ and particle-particle calculations are for $\mathrm{Gap}=4R$.
}
\label{Fig-CR-Cylinder}
\end{figure}

The probability distribution of the contact radiuses is assumed to have an algebraic tail with exponent $\alpha$ and to be translated by $a_0$, that is
\begin{equation}
	P(\bar{a}) 
	\propto 
	\bar{a}^{3\gamma -1} 
	\exp\left[ -\lambda^\alpha \left|\bar{a}^{3}-a_0^3\right|^\alpha  \right]
\label{Eqn-ContactRadiusFit}
\end{equation}
where we have changed variables in \eqref{Eqn-ForceFit} using $\bar{f} \propto \bar{a}^3$. We note that the hertzian relationship $f = K a^3$ does not hold for the average values (i.e., $f_\mathrm{\small av} \ge K a_\mathrm{\small av}^3$) and thus the non-dimensional values are related by the following inequality $\bar{f} \le \bar{a}^3$. The identification of tighter bounds for the inequality and how these bounds can be used to  bound $P(\bar{a})$ are however beyond the scope of this work. Figure \ref{Fig-CR-Cylinder} shows that, as relative density increases, the range of observed contact radiuses is narrower. This transition however occurs faster than in the case of contact forces for particle-particle interactions (cf. Figure \ref{Fig-CF-Cylinder}). In accordance with the behavior of the network of forces, the figure indicates that the distributions, though similar at jamming onset, are very different at full compaction. Figure \ref{Fig-CR-Cylinder} also shows a best fit of the numerical results to equation \eqref{Eqn-ContactRadiusFit} for reference purposes. However, the simplifications proposed for the network of forces, i.e., equations (\ref{Eqn-PP-Coefficients}-\ref{Eqn-PW-Coefficients}), do not hold true for the network of contact radiuses. It is work noting that the choice of $\epsilon_\mathrm{tol}$ does influence the tail of the probability distribution at $\bar{a}=0$ when the system is at jamming onset, see Figure  \ref{Fig-CR-Cylinder}(a). 

%%%%%%%%%
\section{Summary and discussion}
\label{Section-Summary}

We have reported three-dimensional particle mechanics static calculations that predict the microstructure evolution of monodisperse elastic spherical particles during die-compaction up to relative densities close to one. We employ a nonlocal contact formulation that remains predictive at high levels of confinement by removing the classical assumption that contacts between particles are formulated locally as independent pair-interactions. Our approach demonstrates that the coordination number depends on the level of compressibility, i.e., on the Poisson's ratio, of the particles and thus its scaling behavior is not independent of material properties as previously thought. Our results also reveal that distributions of contact forces between particles and between particles and walls, although similar at jamming onset, are very different at full compaction. Both distribution tails transition from exponential to Gaussian under increasing relative density---with particle-wall forces exhibiting an earlier transition. Particle-wall forces are in remarkable agreement with experimental measurements reported in the literature, providing a unifying framework for bridging experimental boundary observations with bulk behavior. The evolution of these microstructural features is not otherwise amenable to direct experimental determination. Furthermore, prediction of such evolution is key to better design, optimize and control many manufacturing processes widely used in pharmaceutical, energy, food, ceramic and metallurgical industries.

We close by pointing out some limitations of our approach and possible avenues for extensions of the analysis.

First, we have generated the initial powder bed by means of a ballistic deposition technique and progressively moving the upper punch until particle rearrangement and small deformations lead to a jammed configuration. This is not the only---perhaps even the best---available protocol for obtaining jammed configurations in rigid containers. A case in point consists in modeling the filling process (see, e.g., \cite{Mateo-2014,Wu-2008}) by solving the equations of motion of the Lagrangian system of particles with a numerical time integrator (see, e.g., \cite{Gonzalez-2010, Gonzalez-2012, Hairer-2006} and references there in). However, this approach, though more realistic, may also result in a jammed backbone and a number of rattlers. The systematic investigation of the effect of the filling protocol on the formation and evolution of the microstructural features studied here at high relative densities is a worthwhile direction of future research.

Second, we have restricted attention to die-compaction of monodisperse system constrained by a rigid cylindrical container and two flat punches.The systematic extension of this analysis to dies and punches of other shapes, hydrostatic compaction conditions, polydispersity and mixtures of particles with different material properties are interesting topics to study.

Third, although our analysis has focused on coordination number, punch and die-wall pressures, and the network of contact forces and radiuses, there are other microstructural features of theoretical and practical importance. For example, it is relevant to study the evolution of material-fabric tensor, force-fabric tensor and pore size distribution during die compaction.

%%%%%%%%%%%%%%%%%%%%%%%%%%%%
\section*{Acknowledgements}
The authors gratefully acknowledge the support received from the NSF ERC grant number EEC-0540855, ERC for Structured Organic Particulate Systems. MG also acknowledges support from Purdue University's startup funds.

%%%%%%%%%%%%%%%%%%%%%%%%%%%%
\bibliographystyle{plainnat}

\end{document}